\documentclass[twocolumn]{aastex62}

\newcommand{\kp}{$K_\mathrm{P}$}		
\newcommand{\vsys}{$V_\mathrm{sys}$}	
\newcommand{\kms}{km\,s$^{-1}$}			
\newcommand{\pname}{HD~209\,458~b}
\newcommand{\water}{H$_2$O}
\newcommand{\hires}{HRCCS}				
\newcommand{\lores}{LRS}				
\newcommand{\cctologl}{CC-to-$\log L$}
\newcommand{\tp}{$T$-$p$}

\received{}
\revised{}
\accepted{}
\submitjournal{ApJ}

\shorttitle{Exoplanet atmospheres at high spectral resolution}
\shortauthors{Brogi \& Line}

\begin{document}


\title{Retrieving Temperatures and Abundances of Exoplanet Atmospheres with High-\\resolution Cross-correlation Spectroscopy}

\correspondingauthor{Matteo Brogi}
\email{m.brogi@warwick.ac.uk}

\author[0000-0002-7704-0153]{Matteo Brogi}
\affiliation{Department of Physics, University of Warwick, Coventry CV4 7AL, UK}
\affiliation{INAF--Osservatorio Astrofisico di Torino, Via Osservatorio 20, I-10025, Pino Torinese, Italy}
\affiliation{Centre for Exoplanets and Habitability, University of Warwick, Gibbet Hill Road, Coventry CV4 7AL, UK}

\author{Michael R. Line}
\affiliation{School of Earth and Space Exploration, Arizona State University, Tempe, AZ 85287, USA}



\begin{abstract}

High-resolution spectroscopy ($R \ge 25,000$) has recently emerged as one of the leading methods for detecting atomic and molecular species in the atmospheres of exoplanets. However, it has so far been lacking a robust method to extract quantitative constraints on the temperature structure and molecular/atomic abundances. In this work, we present a novel Bayesian atmospheric retrieval framework applicable to high-resolution cross-correlation spectroscopy (HRCCS) that relies on the cross-correlation between data and models for extracting the planetary spectral signal. We successfully test the framework on simulated data and show that it can correctly determine Bayesian credibility intervals on atmospheric temperatures and abundances, allowing for a quantitative exploration of the inherent degeneracies. Furthermore, our new framework permits us to trivially combine and explore the synergies between HRCCS and low-resolution spectroscopy (LRS) to maximally leverage the information contained within each. This framework also allows us to quantitatively assess the impact of molecular line opacities at high resolution.  We apply the framework to VLT CRIRES $K$-band spectra of HD 209458 b and HD 189733 b and retrieve abundant carbon monoxide but subsolar abundances for water, which are largely invariant under different model assumptions. This confirms previous analysis of these datasets, but is possibly at odds with detections of \water\ at different wavelengths and spectral resolutions. The framework presented here is the first step toward a true synergy between space observatories and ground-based high-resolution observations. 
\end{abstract}



\section{Introduction}
The field of exoplanet characterization has matured to the point where we can begin to answer fundamental questions regarding planetary climate, composition, and formation, and provide context for understanding our own solar system planets \citep{madhu14, bur14, bai14, cross15}. The community has leveraged the power of ground- and space-based observatories to find and characterize a diverse range of planets, ranging from hot giants to terrestrial-sized, potentially habitable worlds. Atmospheric characterization has emerged as key area of intense focus as of late because the atmosphere is the most readily accessible part of a planet via remote observations.  

The most scientifically valuable measurements of exoplanet atmospheres are those constraining their composition 
and temperature changes with altitude, ideally as a function of orbital phase. The most stringent constraints so far come from observations with low resolution spectroscopy (hereafter \lores, at a resolving power $R = \lambda /\Delta \lambda \le 200$), primarily with the Hubble \citep{evans16,haynes15,knutson14,kreidberg14,mandell13,stevenson14,dem13} and Spitzer Space Telescopes \citep{grillmair08}.  The instruments on board HST permit for near continuous coverage over a broad wavelength range spanning $\sim$0.3-1.7 $\mu$m \citep[e.g.,][]{sing16} split across three pass-bands. In addition, Spitzer provides complementary broadband photometry from 3.5 to 5.4 $\mu$m (and historically out to 8 $\mu$m).  While the coverage is broad, the resolution is very coarse, typically $R \sim$30-200. HST observations in both emission and transmission generally permit order-of-magnitude constraints on the molecular abundance of water \citep[e.g.,][]{Kreidberg2014,lin16,wakeford17,tsiaras18}.  The low resolution and limited near-IR coverage, however, has precluded our ability, for most objects, to sufficiently break degeneracies to constrain (beyond upper limits) the abundances of other key diagnostic molecules like methane, carbon dioxide, carbon monoxide, ammonia, hydrogen cyanide, acetylene.  The James Webb Space Telescope (JWST), anticipated to launch within the next few years, will significantly improve precisions on the aforementioned quantities \citep{greene16}, due to the extremely broad, continuous (0.6-12 \micron), low-to-moderate resolution spectroscopy ($R\sim$ 100-3,500).  However, JWST is going to be a limited resource and it is therefore crucial to pair it with independent but complementary observations.

	A powerful emerging approach for characterizing exoplanet atmospheres is high-resolution cross-correlation spectroscopy (\hires). It leverages the ability of resolving molecular bands into the individual lines and detecting the planet's Doppler shift directly at the $\sim$\kms\ level. In addition, it benefits from the large collective area of ground-based telescopes.    
    Spectral information is extracted through cross-correlation with model templates, which acts as a robust signal filtering technique to recognize the peculiar fingerprint of each species. \hires\ is the best technique so far to unambiguously identify molecules, and it is the only technique to have reliably detected carbon-based molecules in the atmospheres of transiting and non-transiting exoplanets, starting with the pioneering detection of carbon monoxide by \citet{sne10}.
    
    Despite this potential, little work has focused on rigorously determining the {\it abundances} of molecules, the vertical temperature structure, and other fundamental atmospheric properties from \hires\ data. One of the primary challenges in doing so is placing \hires\ within a robust atmospheric retrieval framework. Contrarily to \lores\ data, where spectra are either calibrated in flux or measured in comparison to a reference star, \hires\ data is ``self-calibrated'', meaning that the broad-band information and the time variations of flux at each wavelength are divided out of the data. With such normalization, the small variations due to the planet atmosphere are all measured {\sl relative} to the stellar flux. This peculiarity, together with small residual broad-band structure due to imperfect normalization, results in the loss of a reliable continuum which inhibits the use of the standard (data spectrum -- model spectrum) residual vector used in typical \lores\ parameter estimation via chi-square. Cross-correlating with model spectra, besides providing a matching filter for robust identification of species, is also insensitive to any residual broad-band variations. 
    
    The challenge with retrieving atmospheric parameters from high-resolution spectra is converting the cross-correlation values into a goodness-of-fit estimator. \citet{bro12} developed a statistical test to assess the significance of cross-correlation signals by comparing the distribution of cross-correlation values around the planet radial velocity (typically labeled as the ``in-trail'' sample) to the values away from it (the ``out-of-trail'' sample). This strategy has been widely adopted in the literature since then \citep{bir13, bir17, bro13, bro14, bro18, nugroho17, hawker18}. Due to the necessity to compute a statistically significant number of cross-correlation values (typically thousands of values mapping the planet's orbital and systemic velocity), its application is limited to the evaluation of a relatively small ($\sim100$) number of models. Furthermore, this test is somewhat sensitive to the range around the planet radial velocity chosen for building the {\sl in-trail} distribution.
    
    \citet{bro17} introduce a different method to overcome some of the above limitations. They estimate as accurately as possible the {\sl model} cross-correlation function and compare it to the {\sl observed} cross-correlation function via chi-square fitting. To replicate as closely as possible astrophysical, instrumental, and data analysis effects, each model is added to the real data, albeit at a sufficiently low level so that the noise properties and data analysis are not altered. This alternative strategy still requires significant computational resources and is therefore limited to the evaluation of only a few thousands model \hires\ spectra sampled from a pre-existing \lores\ posterior. Such sparse sampling is only sufficient to constrain confidence intervals within the 3-$\sigma$ level, and on a limited portion of the parameter space. As a consequence, full independent retrievals on \hires\ data cannot be performed. Conditioning the \hires\ retrieval on the \lores\ retrieval implicitly assumes that the two datasets contain the same amount of evidence (or weighting), which is not necessarily the case given the different spectral range and signal to noise of the observations. Lastly, since neighboring cross-correlation values can be correlated (depending on the sampling in velocity and the instrumental resolution), chi-square is not necessarily the correct statistic to compare cross-correlation functions.
    
    The primary goal of this paper is to introduce a robust and unbiased framework to perform Bayesian retrieval analyses on \hires\ data, free from the limitations of previous approaches. In Section~\ref{sec:framework} we define a likelihood function for high-resolution spectra, and we describe its implementation into a nested sampling Bayesian estimator. In Section~\ref{sec:testing} we describe the setup used to test our new Bayesian framework, including a simulated dataset replicating previous work on dayside spectroscopy of \pname. We present the analysis of the simulated dataset (Section~\ref{sec:analysis}), the excellent match between the retrieved and the modeled atmospheric parameters (Section~\ref{sec:retrieval}), the effects of uncertainties in the line list of water vapor (Section~\ref{sec:h2o_xsec}) and the increased precision when combining space and ground observations (Section~\ref{sec:lds_hds_combi}). We then apply the framework to real observations of two exoplanets, HD~189\,733~b and \pname, in Section~\ref{sec:real_data}. We conclude in Section~\ref{sec:conclusions} by highlighting future applications of this framework, in particular coordinated observations with space and ground telescopes.
	
\begin{figure}
\plotone{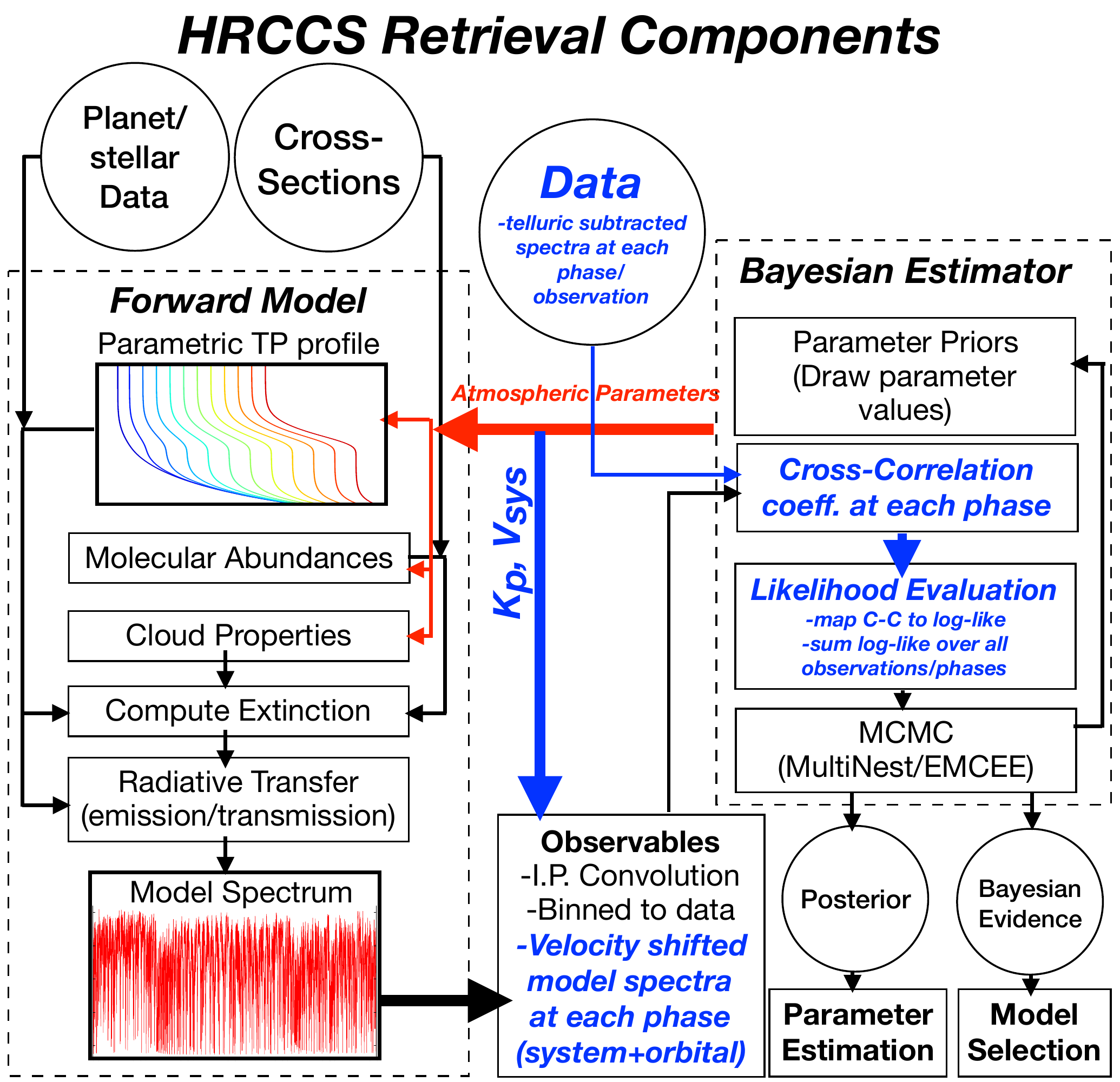}
\caption{Illustration of the key components of the \hires\ retrieval algorithm. Many components are the same as in a classic ``low-resolution'' retrieval: the forward model, the data, and the Bayesian Estimator.  The key novelties (indicated in blue) required to perform retrievals in cross-correlation space are the inclusion of the radial velocities (system+orbital), a telluric subtracted ``data-cube'' of normalized spectrum vs. time/phase, and a mathematical mapping from correlation-coefficient to a likelihood function that can be used inside a parameter estimation package such as an MCMC \citep[Figure modified from][with permission]{MacMad2017}.}
\label{fig:retrieval_flow}
\end{figure}	

\section{A Novel HRCCS Atmospheric Retrieval Framework}\label{sec:framework}

At its core, an HRCCS retrieval is no different than the typical \lores\ retrievals applied in numerous previous works \citep[see, e.g., review by][]{madhu18}.  The key components of any retrieval algorithm are the data or observable, the forward model which maps the quantities of interest onto the data/observable, and the Bayesian estimator that optimizes the forward model parameters (or range thereof) given a likelihood function (Figure~\ref{fig:retrieval_flow}). Below we describe each of these key components and how they are adapted within our \hires\ retrieval framework.

\subsection{The Forward Model}\label{sec:modeling}
A typical spectral retrieval forward model is a radiative transfer routine that takes in gas abundances, cloud properties, vertical thermal structure information, and/or geometric information to produce a transmission or emission spectrum.  In this work we leverage the CHIMERA forward model \citep{lin13a,lin13b,lin14,lin15,lin16,lin17}.  The CHIMERA\footnote{A version of the transmission spectrum code is publically avaialable through STScI's ExoCTK package: https://github.com/ExoCTK/chimera} forward model offers the flexibility to handle a variety of atmospheric assumptions from simple 1D non-self-consistent \citep{lin13a,lin14,lin16} to full 1D self-consistent thermo-chemical radiative-convective equilibrium \citep{Arcangeli2018,Kreidberg2018,Mansfield2018} to quasi 2D/3D parameterizations and observing geometries \citep{Feng2016,LineParm2016} as well as flexible treatments of opacities (e.g., line-by-line, line-sampling, and correlated-$k$) based upon a pre-tabulated line-by-line ($<$0.01 cm $^{-1}$) absorption cross section database \citep{Freedman2008, Freedman2014}. 

Specifically, here, we explore the classic {\sl free} and {\sl chemically-consistent} versions described in \cite{kreidberg2015} but in the emission geometry. In the {\sl free} forward model we include as free parameters the H$_2$O and CO mixing ratios (constant with pressure), H$_2$-H$_2$/He collision induced absorption-CIA, and the 3-parameter\footnote{$\log(\kappa_\mathrm{ir})$ = log of Planck mean infrared opacity; $\log(\gamma)$ = log of the visible-to-infrared opacity; $T_\mathrm{irr}$ = top-of-atmosphere irradiation temperature.} disk-integrated temperature-pressure profile (\tp~profile) parameterization of \citet{Guillot2010}\footnote{Different temperature profile parameterizations can be explored in a future investigation.}.  In the {\sl chemically consistent} forward model we utilize the same \tp~profile parameterization but parameterize the composition with a metallicity ([M/H], where solar is 0) and log-carbon-to-oxygen ratio ($\log$(C/O), where solar is $-0.26$) under the assumption of pure thermochemical equilibrium computed using the NASA CEA2 routine \citep{Gordon1994,kreidberg2015} by scaling the solar elemental abundances of \citet{Lodders2009}. In forward modeling used in this work we include as opacities  H$_2$O, CH$_4$, CO, CO$_2$, NH$_3$, HCN, and H$_2$-H$_2$/He CIA, however for nearly solar abundances only H$_2$O and CO present themselves over the $K$-band spectrum.  The output from the forward model is a line-by-line top-of-atmosphere flux over the $K$-band (2.26--2.35 \micron) which is then normalized to the stellar spectrum and planet-to-star area ratio. 

The forward model spectra are then Doppler-shifted by spline interpolation\footnote{with python's {\tt  scipy.interpolation} package} \citep{bro14} based on the semi-amplitude of the planet radial velocity (\kp), the systemic velocity (\vsys), and the barycentric velocity of the observer $V_\mathrm{bary}$, according to:
\begin{equation}\label{eq:pl_rv}
V_\mathrm{P}(t) = V_\mathrm{sys} + V_\mathrm{bary}(t) + K_\mathrm{P}\sin\left[2\pi\varphi(t)\right],
\end{equation}
where we have neglected cross-terms in velocity due to their small impact (on the order of m s$^{-1}$). The Doppler-shifted model spectra are finally convolved with the appropriate instrumental profile. Due to the fact that velocities in Eq.~\ref{eq:pl_rv} are not perfectly known, we introduce two additional key parameters: a differential systemic velocity (d\vsys) and a differential planet radial velocity (d\kp). With typical uncertainties of a few \kms, detectable at the spectral resolution and signal-to-noise ratio of \hires\ observations, neglecting these differential velocities would lead to incorrect localization of the planet's signal.

\subsection{The Bayesian Estimator: Cross-Correlation to Log-Likelihood Mapping}

Bayesian estimation necessarily requires a well defined likelihood function \citep{gregory2005, sivia2006,DFM2013,Feroz2009}. The standard likelihood function in \lores\ analysis is a Gaussian, or similarly a quadratic or chi-square log-likelihood. The challenge with \hires\ data is to exploit the power of the large number of spectral lines and relatively well known orbital velocity via the cross-correlation between a template model and the phase dependent data. Once a mapping from the readily calculable cross-correlation coefficient to log-likelihood is found, one can employ the standard suite of Bayesian analysis tools including parameter estimation, prior inclusion, and model selection. We utilize the powerful {\tt pymultinest} tool \citep{Buchner2014}, a python wrapper to {\tt multinest} \citep{Feroz2009}, to perform all of the parameter estimation within our HRCCS retrieval framework.   

In building this \hires\ likelihood function, we want to exploit some unique characteristics of the analysis of high resolution spectra. Firstly, the continuum is divided out of the data to allow for their self-calibration (i.e. to eliminate the necessity for a reference star). Secondly, the planetary signal is extracted via cross-correlation with model spectra, so our metric for the goodness of fit must incorporate the cross-correlation function or a closely related quantity. Thirdly, the sign of the correlation-coefficient matters, as it allows us to discriminate between emission lines incorrectly fitted with an absorption spectrum or vice-versa. Lastly, although the cross-correlation function is by definition normalized and thus insensitive to scaling of the model and/or the data, observations have a finite signal-to-noise ratio (SNR). This means that two nearly identical model spectra in terms of line-to-line ratios, but differing by orders of magnitude in the overall line strength should yield a different likelihood. 

\subsubsection{Previous Cross-Correlation to Log-Likelihood Mappings}\label{sec:past_map}
The idea of mapping the correlation coefficient to a log-likelihood (hereafter \cctologl) is not new. One of the earlier applications of such a mapping in the exoplanet community was in the context of precise stellar radial velocities \citep{zuc03}. The aim of \citet{zuc03} was to derive a formalism that facilitated combining radial velocity measurements obtained via cross-correlation from multiple spectral regions and at varying SNR. Starting from the definition of $\chi^2$, \citet{zuc03} derived the following \cctologl\ mapping: 
\begin{equation}\label{eq:Zucker_Map}
\log(L) = -\frac{N}{2}\log(1-C^2). 
\end{equation}
The key feature of this mapping is that the $\log L$ depends on the square of correlation coefficient ($C$).  For our application, the use of the square of the cross-correlation function limits the sensitivity to inversion layers. Although the actual line shape in day-side spectra can be quite complex and it is not a trivial 
``sign flip" when going from non-inverted to inverted \citep[e.g.,][]{schwarz15}, spectra with emission lines cross correlated with absorption models (and vice-versa) produce some anti-correlation that would not be discriminated due to the insensitivity to the sign of $C$. In addition, the log-likelihood of \citet{zuc03} weights the spectra based on their cross-correlation value, assuming therefore that each spectrum delivers a peak significantly above the level of the noise. Although this is appropriate for stellar radial velocities, it is not the case for planetary radial velocities, and we thus expect this likelihood to struggle at low SNR levels.  This mapping should work well for high-signal to noise spectra of isolated objects like brown dwarfs and directly imaged planets, as demonstrated in \citet{bowler17}.

\citet{loc14} were the first to apply a \cctologl\ mapping to high-resolution exoplanet spectroscopy of the non-transiting planet $\tau$~Bo\"otis b. Their analysis comprises two steps. Firstly, the formalism of \citet{zuc03} as implemented in the two-dimensional routine TODCOR is used to combine all the cross-correlation functions taken at a certain epoch (e.g. during one night of observations) into a maximum-likelihood estimator, i.e. an effective cross-correlation value. Subsequently, a \cctologl\ mapping is derived as:
\begin{equation}\label{eq:Lockwood_Map}
\log(L) = C
\end{equation}
and used to combine CCFs taken at different epochs. An important underlying hypothesis of their formalism, which is instead violated by our CRIRES observations, is that the change in planet radial velocity during one set of observations is negligible with respect to the instrumental resolution. This is ensured by typically taking NIRSPEC spectra when the planet is in quadrature. An additional substantial difference with our analysis is that the cross-correlation function contains both the stellar and the planet spectrum, i.e. it is a two-dimensional cross correlation with the stellar coefficients dominating the planet coefficients by orders of magnitude. 

The mapping of \citet{loc14} was also recently adopted in \citet{pis18} to combine Keck NIRSPEC $K$-band data with Spitzer for the transiting hot-Jupiter Kelt-2~Ab. However, we again stress that such formalism cannot be applied to our data analysis where we make use of the change in planet radial velocity with time.
We show indeed in Section \ref{sec:logL_distr} that if we incorrectly apply Equation~\ref{eq:Lockwood_Map} to our data, the resulting $\log L$ is not distributed as a $\chi^2$ (Figure~\ref{fig:wilks_test}) as Wilks' Theorem demands \citep{wilks1938}.

\subsubsection{A New Mapping}\label{sec:newlogL}
In this Section we derive a new \cctologl\ mapping that leverages all of the aforementioned key components of a \hires\ observation. The starting point for building our mapping closely matches the derivation in section 2 of \citet{zuc03}.
We denote with $f(n)$ a single observed spectrum, where $n$ is the bin number, or spectral channel.

We compute a template spectrum $g(n)$ in the same reference frame as the data. We assume that the model describing the data is:
\begin{equation}
f(n) = a\,g(n-s) + d_n,
\end{equation}
where $a$ is a scaling factor, $s$ a bin/wavelength shift, and $d_n$ the noise at bin $n$. 

It is important that $f(n)$ and $g(n)$ are continuum subtracted. In our current analysis, and following common implementation of numerical cross-correlation routines, we achieve this by subtracting the mean from each of the vectors prior to cross-correlation. Under these assumptions we have that $\sum_nf(n)=0$ and $\sum_ng(n)=0$.

We assume that the noise is Gaussian distributed at each pixel with standard deviation $\sigma$. The Likelihood function $L$ of our model is:
\begin{eqnarray}
\nonumber L & = & \prod_n \frac{1}{\sqrt{2\pi\sigma^2}} \exp\left\{-\frac{[f(n) - a\,g(n-s)]^2}{2\sigma^2}\right\} = \\
& = & \left( \frac{1}{\sqrt{2\pi\sigma^2}} \right)^N \exp\left\{-\sum_n \frac{[f(n) - a\,g(n-s)]^2}{2\sigma^2}\right\},
\end{eqnarray}
where $N$ is the total number of spectral channels. This is typically the number of pixels per spectrum (or per detector, or per spectral order). The log-likelihood can be derived from the above:
\begin{equation}\label{eq:logL_generic}
\log(L) = -N\log\sigma - \frac{1}{2\sigma^2} \sum_n [f(n) - a\,g(n-s)]^2,
\end{equation}
after neglecting constant additive terms. We note that throughout this Section the function ``$\log$'' will indicate the natural logarithm.

At this stage our analysis diverges from \citet{zuc03}. We impose that the scaling factor is unity ($a=1$). Physically, this means that we want the overall strength of spectral lines in the model to match the observed data. Practically, since our data $f(n)$ is normalized and telluric corrected, any residual variations is relative to the stellar continuum. We therefore scale $g(n)$ by the stellar flux and planet-to-stellar area ratio (equation 12 below). We note that the formalism in \citet{zuc03} does not specify whether the cross correlation is performed with a model spectrum, a binary mask, or generically a template. Therefore, $a$ does not carry any physical meaning in that context, as the units of the template are arbitrary. This is why in their work $a$ is substituted with its maximum-likelihood estimator $\hat{a}$.


Eq.~\ref{eq:logL_generic} contains another variable, $\sigma$. We do not fix this, as it is difficult to estimate the exact level of noise in each spectral channel, especially under imperfect removal of the telluric spectrum. We instead compute the maximum likelihood estimator $\hat{\sigma}$ by nulling the partial derivative of the log-likelihood:
\begin{eqnarray}
\nonumber 0 & = & \frac{\partial\log(L)}{\partial \sigma} \\
\nonumber \frac{N}{\hat\sigma} & = & \frac{1}{\hat{\sigma}^3} \sum_n [f(n) -g(n-s)]^2 \\
\hat{\sigma}^2 & = & \frac{1}{N}\sum_n [f(n)-g(n-s)]^2
\end{eqnarray}
Substituting $\hat{\sigma}$ into $\log(L)$ and neglecting constant additive terms we get:
\begin{eqnarray}
\nonumber \log(L)  =  -N\log\left\{ \sqrt{ \frac{1}{N}\sum_n [f(n)-g(n-s)]^2} \right\} - \frac{N}{2} \\
\nonumber  =  -\frac{N}{2}\log\left\{ \frac{1}{N}\sum_n [f(n)-g(n-s)]^2 \right\} \nonumber \\
\nonumber  =  -\frac{N}{2}\log\left\{ \frac{1}{N}\sum_n [ f(n)^2 - 2f(n)g(n-s) + g(n-s)^2] \right\} \\
&&
\end{eqnarray}
We can now write the formulas for the variance of the data ($s_f^2$), the variance of the model ($s_g^2$), and the cross-covariance $R(s)$:
\begin{eqnarray}
\nonumber s_f^2 & = & \frac{1}{N}\sum_n f^2(n) \\
\nonumber s_g^2 & = & \frac{1}{N}\sum_n g^2(n-s) \\
\nonumber R(s) & = & \frac{1}{N}\sum_n f(n)g(n-s)
\end{eqnarray}\label{eq:our_mapping}
Substituting them into $\log(L)$ leads us to:
\begin{equation}\label{eq:logL_CC2}
\log(L) = -\frac{N}{2}\log\left[ s_f^2 -2R(s) + s_g^2 \right]
\end{equation}
Factorizing out the product $s_fs_g$ we can make the cross-correlation appear:
\begin{eqnarray}\label{eq:logL_CC3}
\nonumber \log(L) & = & -\frac{N}{2} \left\{ \log{(s_fs_g)} + \log \left[ \frac{s_f}{s_g}+\frac{s_g}{s_f}-2 \frac{R(s)}{\sqrt{s_f^2s_g^2}} \right] \right\} = \\
& = & -\frac{N}{2} \left\{ \log{(s_fs_g)} + \log \left[ \frac{s_f}{s_g}+\frac{s_g}{s_f}-2C(s)\right] \right\}
\end{eqnarray}
with the correlation coefficient
\begin{eqnarray}
C(s) = \frac{R(s)}{\sqrt{s_f^2s_g^2}}
\end{eqnarray}
For practical applications, Eq.~\ref{eq:logL_CC2} is slightly faster to compute than Eq.~\ref{eq:logL_CC3} and is the preferred choice for our numerical implementation. It is important to note that the data variance $(s_f^2)$ only needs to be computed once at the end stage of the data analysis (step 7 in Section \ref{sec:analysis} below). However $s_g^2$ will change as function of the model tested, and also to a lesser extent as function of the Doppler shift tested. Therefore in our analysis we will recompute $s_g^2$ every time a model is evaluated, and for each of the spectra in the time sequence. 

Eq.~\ref{eq:logL_CC2} preserves the sign of the cross-covariance, and will therefore discriminate between correlation and anti-correlation. This is a direct consequence of imposing $a=1$. In addition, when the variance of the data and the (scaled) model differ significantly, the likelihood decreases accordingly. This incorporates a metric for comparing the average line depth to the SNR of the data. 

It is important to realize that if we carried on the mathematical calculations with the scaling factor $a$ as an explicit variable, and then imposed $\partial\log(L) / \partial a = 0$ at the stage of Equation~\ref{eq:logL_CC2} (had we kept an $a$, and $a^2$ multiplier in front of the $R(s)$ and $s_g^2$ terms, respectively), we would have obtained as solution $a=1$. This means that our physically-motivated choice of $a=1$ also corresponds to choosing the maximum-likelihood estimator for this variable.

\begin{table*}
	\centering
	\caption{Forward model parameters and uniform prior ranges. The first 3 parameters are required to Doppler shift and scale the template model spectrum. The second 3 control the temperature-pressure profile. The planet metallicity and carbon-to-oxygen ratio replace the \water\ and CO mixing ratios when using the {\it chemically consistent} model. In most tests, the {\it free} model includes only H$_2$O, CO, and H$_2$/He CIA as opacity sources whereas the {\it chemically consistent} model includes opacities from H$_2$O, CH$_4$, CO, CO$_2$, NH$_3$, HCN, and H$_2$-H$_2$/He CIA. Each model ({\it free} or {\it chemically consistent}) has 8 total free parameters unless specified otherwise.\label{tab:model_params}}
	\begin{tabular}{lccc} 
		\hline
		Parameter & Symbol & Uniform Prior Range & Model Input ``Truth'' \\
        \hline
        {\it Doppler/Scale Parameters}  &   &  & \\
        Relative Systemic Velocity & d$V_\mathrm{sys}$ & $-50$ - 50 \kms & 0   \\
        Relative Planet Radial Velocity & d\kp & $-50$ - 50 \kms & 0   \\
        Model Scale Factor & log($a$) & $-2$ - 2  & 0.0 \\
        \hline
        {\it Temperature-Pressure Profile Parameters}  &   &  & \\
        Planck Mean IR Opacity & log($\kappa_\mathrm{ir}$) & $-3$ - 1 & $-1$ \\
        Visible to IR Opacity & log($\gamma$) & $-4$ - 2 & $-1.5$ \\
        Irradiation Temperature & T$_\mathrm{irr}$ & 300 - 2800 K & 1400  \\
        \hline
        {\it Free Retrieval Abundance Parameters}  &   &  & \\
        H$_2$O Mixing Ratio & log(H$_2$O) & $-12$ - 0 & $-3.4$    \\
        CO Mixing Ratio & log(CO) &  $-12$ - 0 & $-3.22$  \\
        \hline
        {\it Chemically Consistent Abundance Parameters}  &   &  & \\
        Metallicity & $[M/H]$ & $-2$ - 2 & 0.0  \\
        Carbon-to-Oxygen Ratio & log(C/O) & $-2$ - 1 & $-0.26$   \\

	\end{tabular}
\end{table*}
All the quantities listed in Eq.~\ref{eq:logL_CC2} are obtained as byproducts of the current analysis techniques of high-resolution spectra. In Section \ref{sec:analysis} we discuss additional details of the data analysis important for the application of this formalism.



\begin{deluxetable*}{lcCcCr}
	\tablecaption{Relevant parameters for the systems HD 209\,458 and HD 189\,733 used throughout the paper.  Parameter ranges the phase range and radial velocity for HD 209458 system are given for both nights. References are K07 = \citet{knutson07}, T08 = \citet{torres08}, T09 = \citet{triaud09}, A10 = 	\citet{agol10}, S10=\citet{southworth10} B16 = \citet{bro16}, B17 = \citet{bro17}.\label{tab:sys_pars}}
	\tablecolumns{6}
	\tablehead{
	\colhead{} & \colhead{} & \multicolumn{2}{c}{HD 209\,458} & \multicolumn{2}{c}{HD 189\,733} \\
	\colhead{Parameter} & \colhead{Symbol} & \colhead{Value} & \colhead{Reference} 
	& \colhead{Value} & \colhead{Reference} }
	\startdata
	Stellar radius (R$_{sun}$) & R$_\star$ & 1.162 & S10 & 0.756 & T08 \\
	Effective temperature (K) & $T_\mathrm{eff}$ & 6065 & T08 & 5040 & T08 \\
	\hline
	Planet Gravity (log$_{10}$, cgs) & log($g_p$) & 2.96 & S10 & 3.28 & \nodata\\
	Planet Radius (R$_J$) & $R_\mathrm{P}$ & 1.38 & S10 & 1.178 & T09 \\
	Radial-velocity amplitude (\kms) & \kp & 145.9  & B17 & 152.5  & B16 \\
	Phase Range (\# spectra) & $\phi$ & [0.506, 0.578]\ (59)  & K07 & [0.383, 0.479]\ (110) & A10 \\
	&  & [0.557, 0.622]\ (54)  & K07 & \nodata & \nodata \\
	Radial Velocity (km/s) & $V_{sys}+V_{bary}$ & [-26.92, -26.25] 
	& K07+T08 & [-9.40, -8.84] & T09+A10 \\
	&  & [-13.44, -12.81]  & K07+T08 & \nodata & \nodata \\
	\enddata
\end{deluxetable*}

\section{Tests on Simulated Data}\label{sec:testing}
In this section we demonstrate, on a simulated emission spectrum dataset, the feasibility and utility of our novel \hires\ retrieval framework and \cctologl\ mapping presented in Section \ref{sec:framework}.  We start by describing the construction of the simulated dataset in Section \ref{sec:sim_dataset} and its analysis in Section \ref{sec:analysis}. We present the ``fiducial'' retrieved constraints in Section \ref{sec:retrieval}, compare the constraints derived from different \cctologl\ mappings in Section \ref{sec:logL_distr}, explore the impact of different water line-lists in Section \ref{sec:h2o_xsec}, and finally combine in a coherent way the high-resolution spectra with a simulated Hubble Space Telescope Wide Field Camera 3 (HST WFC3) \lores\ dataset in Section \ref{sec:lds_hds_combi}.   

\subsection{Construction of the Simulated Dataset}\label{sec:sim_dataset}

One half night of data is simulated based on real CRIRES observations of \pname\ \citep{schwarz15, bro17}.  The synthetic data-set incorporates the photon noise from the star, variations in the Earth's transmission spectrum with airmass,  variable detector efficiency, the phase dependent Doppler shift of the planet, and the time-dependent instrumental profile. This simulated dataset constitutes the basis to test the retrieval framework presented in the previous Sections, as it incorporates all the major sources of uncertainties in the analysis of \hires\ data. 

\begin{figure}
	\includegraphics[width=8.5cm]{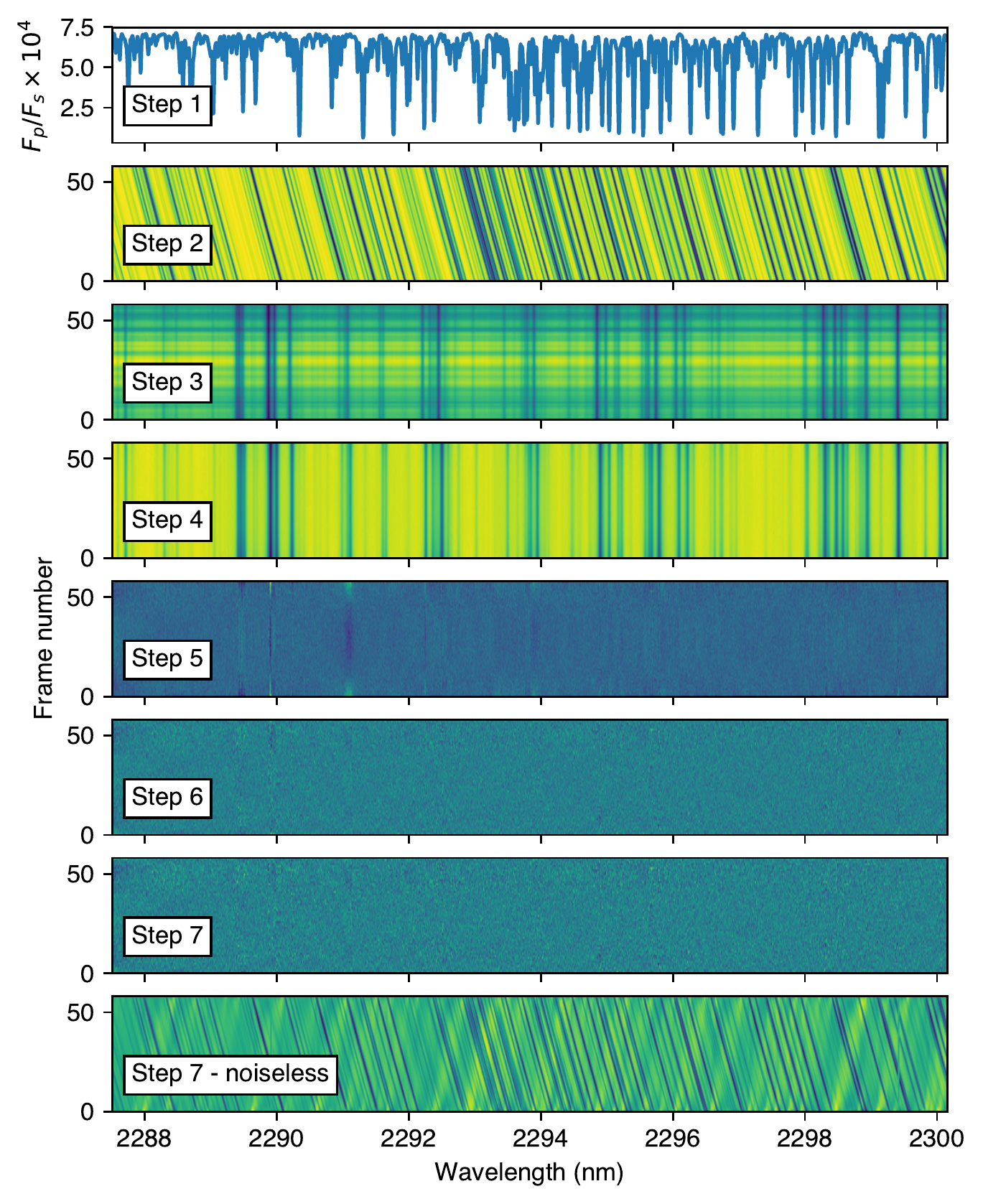}
    \caption{Step-by-step visualization of the process used to construct a simulated dataset (steps 1-3, detailed in Section \ref{sec:sim_dataset}) and analyze it (steps 4-7, detailed in Section \ref{sec:analysis}). This sequence is based on real on-sky performances of the infrared spectrograph CRIRES while observing the exoplanet HD 209\,458 b. Shown is the spectral sequence computed for first detector of the spectrograph (out of four detectors), incorporating all the major astrophysical and instrumental sources of noise. This simulated dataset is used to test the Bayesian framework explained in Section \ref{sec:framework}. The analysis at Steps 4-7 is also performed on the observed datasets, as explained in Section \ref{sec:real_data}}
    \label{fig:demo_flow}
\end{figure}

To generate this dataset, we compute a solar-composition planet atmosphere with parameters listed in Table~\ref{tab:model_params}, and using the modeling tools described in Section \ref{sec:modeling}. We run the computations over the wavelength range 2267-2350 nm (matching the CRIRES setup of the real observations) at a resolving power of $R\sim440,000$ and scale the model to the stellar flux of HD~209\,458 via:
\begin{equation}\label{eq:scaling}
F_\mathrm{scaled}(\lambda) = \frac{F}{B(\lambda, T_\mathrm{eff})} \left( \frac{R_\mathrm{P}}{R_\star} \right)^2
\end{equation}
where $F$ is the model flux in W~m$^{-2}$~m$^{-1}$, $R_\mathrm{P}$ and $R_\star$ are the stellar and planet radii respectively, and $B$ the Planck function at the stellar effective temperature $T_\mathrm{eff}$ approximating the stellar spectrum (Table~\ref{tab:sys_pars}). The top panel (Step 1) of Figure~\ref{fig:demo_flow} shows a small portion of this spectrum in the wavelength range corresponding to detector 1 of CRIRES.

We adopt a Keplerian circular velocity of 145.9 \kms, i.e. the literature value for \pname, and a combination of systemic and barycentric velocity to match the actual observations of night 1 in \citet{bro17} (Table \ref{tab:sys_pars}). The scaled model $F_\mathrm{scaled}$ is Doppler-shifted according to the radial velocity at each epoch of observations computed via Eq.~\ref{eq:pl_rv}, and saved in a matrix $F^\prime(\lambda,t)$.  In this test case, the observations span 1024 pixels/wavelength channels and 59 separate integrations (spectra) covering phases 0.506 - 0.577 resulting in $\sim75$ \kms\ change in Doppler shift throughout the sequence (Figure \ref{fig:demo_flow}, Step 2).

The wavelength- and time-dependent transparency $T(\lambda, t)$ of the Earth's atmosphere (the telluric spectrum) is computed via the ESO Skycalc command-line tool which is based on the Cerro Paranal Sky Model \citep{nol12, jon13}. The model takes into account the sky position of the target at the time of the observation and meteorological data, except for the precipitable water vapor (PWV) that needs to be adjusted manually. We find a good match to the HD 209\,458 dataset by adopting the average PWV of 2.5 mm for Cerro Paranal. 

We measure the average flux levels in the observed CRIRES spectra by taking the median of their brightest pixels, and we compute the number of recorded photoelectrons by multiplying by the exposure time and detector gain of those observations. This gives us a vector $\epsilon(t)$, where we neglect any wavelength dependence of the instrumental throughput. Such dependence certainly occurs even over the small wavelength range of one CRIRES detector (10-15 nm) and indeed measurable trends in the continuum (typically a slope) are on the order of $\sim$1\%. However, owing to the good level of thermal stability of CRIRES (0.01 K over a half night), these effects are not time-dependent and are therefore divided out during telluric removal.

The noiseless modeled dataset is obtained by combining all the above sources:
\begin{equation}\label{eq:noiseless_model}
F_\mathrm{mod}(\lambda,t_i) = \left[ 1 + F^\prime(\lambda,t_i) \right] \,T(\lambda,t_i)\,\epsilon(t_i), 
\end{equation}
where the product is scalar, i.e. computed element by element, and adding 1 accounts for the normalized stellar spectrum. 

In the bright source limit, the noise budget is completely dominated by the stellar photons, so the noise matrix is governed by Poisson statistics. This leads to the following noise matrix:
\begin{equation}\label{eq:noise_matrix}
N_\mathrm{mod}(\lambda,t) = \mathcal{N}(0,1)\,\sqrt{F_\mathrm{mod}(\lambda,t)},
\end{equation}
where $\mathcal{N}(0,1)$ is a Normally-distributed random variable.
The simulated spectrum is the sum of the recorded photons and the noise matrix, i.e.
\begin{equation}\label{eq:noisy_model}
F_\mathrm{sim}(\lambda,t) = F_\mathrm{mod}(\lambda,t) + N_\mathrm{mod}(\lambda,t).
\end{equation}

The panel labeled with Step 3 in Figure~\ref{fig:demo_flow} shows the final simulated ``raw data'' product.  The obvious features seen in the final data matrix are the stationary telluric lines (dark vertical stripes) and the airmass-throughput dependence (horizontal stripes). This synthetic dataset is then processed using the procedure outlined below. 

\subsection{Analysis of the simulated dataset}\label{sec:analysis}
The analysis devised for processing data within this new dataset is largely adapted from previous literature, however there are some caveats related to the nature of our cross-correlation-to-likelihood mapping that require particular care. In Figure \ref{fig:demo_flow}, this analysis is labeled with Steps 4-7, which we detail below:
\begin{itemize}
\item Step 4: Each spectrum is calibrated in wavelength by comparing the pixel position of telluric lines in the observed spectra to their theoretical wavelength obtained from a telluric model. As in \citet{flowers2018}, a common wavelength solution with constant space in velocity (constant $d\lambda / \lambda$) is computed and data are re-gridded to this solution via spline interpolation. Each spectrum is normalized by the median of its brightest 300 pixels to correct for throughput variations. Re-gridding is necessary for estimating the instrumental profile (IP) of CRIRES, which is done at this stage through the procedure described in \citet{ruc99} and implemented in \citet{bro16} and \citet{flowers2018}. The IP is used to convolve the model spectra at a later stage. 
\item Step 5: The spectra are averaged in time and the mean spectrum is fitted to each observed spectrum with a second order polynomial. This procedure removes the main variability in the depth of methane telluric lines, however residuals are still visible at the position of water vapor telluric lines, which behave differently from methane due to the different scale height in the Earth's atmosphere and temporal changes in humidity.
\item Step 6: These extra changes in water telluric lines are corrected by modeling the flux in each spectral channel as function of time with a second order polynomial, and dividing out the fit.  Since the planet's orbital motion produces a time-varying Doppler shift of the spectrum, spectral lines from the exoplanet's atmosphere will shift across adjacent spectral channels and will be nearly unaffected by the above correction. 
\item Step 7: Any further alteration of the noise properties in the data must be avoided, because they would alter the data variance $s_f^2$ in equation~\ref{eq:logL_CC2}. Consequently, the common practice of weighting down noisy spectral channels by dividing them through their variance in time must be avoided. In this revised analysis, noisy columns above 3$\times$ the standard deviation of the matrix are masked and not used at the cross-correlation stage. The total number of channels $N$ is modified accordingly for subsequent use in equation~\ref{eq:logL_CC2}.
\end{itemize}

As in previous work, the above analysis exploits the fact that the planet spectrum is subject to a variable Doppler shift during a few hours of observations, whereas the contaminant signals (telluric and stellar) are stationary or quasi-stationary. However, we note that steps 5 and 6 are achieved with a variety of methods in the literature, either by fitting airmass dependence and then sampling time variations in common between spectral channels directly from the data \citep{bro12,bro13,bro14,bro16}, or by applying blind de-trending algorithms such as principal component analysis \citep{dek13, pis16, pis17} or Sysrem \citep{bir13, bir17}. All these approaches rely on the assumption that a certain spectral line from the exoplanet spectrum sweeps several spectral channels (several columns in our data matrix) during one night of observations, thus not influencing the detrending process significantly. This assumption is only true in first approximation. In reality, whatever algorithm is applied to the data, the planet signal is stretched and scaled in the process. We show an example of this alteration in the bottom panel (Step 7 noiseless) of Figure~\ref{fig:demo_flow}, which is the end point of the data analysis applied to a noiseless dataset. When compared to the initial spectrum at Step 2, the end result clearly shows artifacts and scaling effects due to telluric removal. If unaccounted for, these artifacts can bias the retrieved planet parameters (velocities, abundances, and thermal vertical structure).

As in previous work, the data at Step 7 is cross correlated with models, and each cross-correlation value is converted into $\log(L)$ value through Eq.~\ref{eq:logL_CC2}. To account for the stretching of the planet signal, we repeat Steps 1-7 on each of the tested model spectra, but without adding the noise matrix $N_\mathrm{mod}$. This mimics the effects of the data analysis on the model, and eliminates the biases, at a small computational cost. 

\begin{figure}
	\includegraphics[width=1.1\columnwidth]{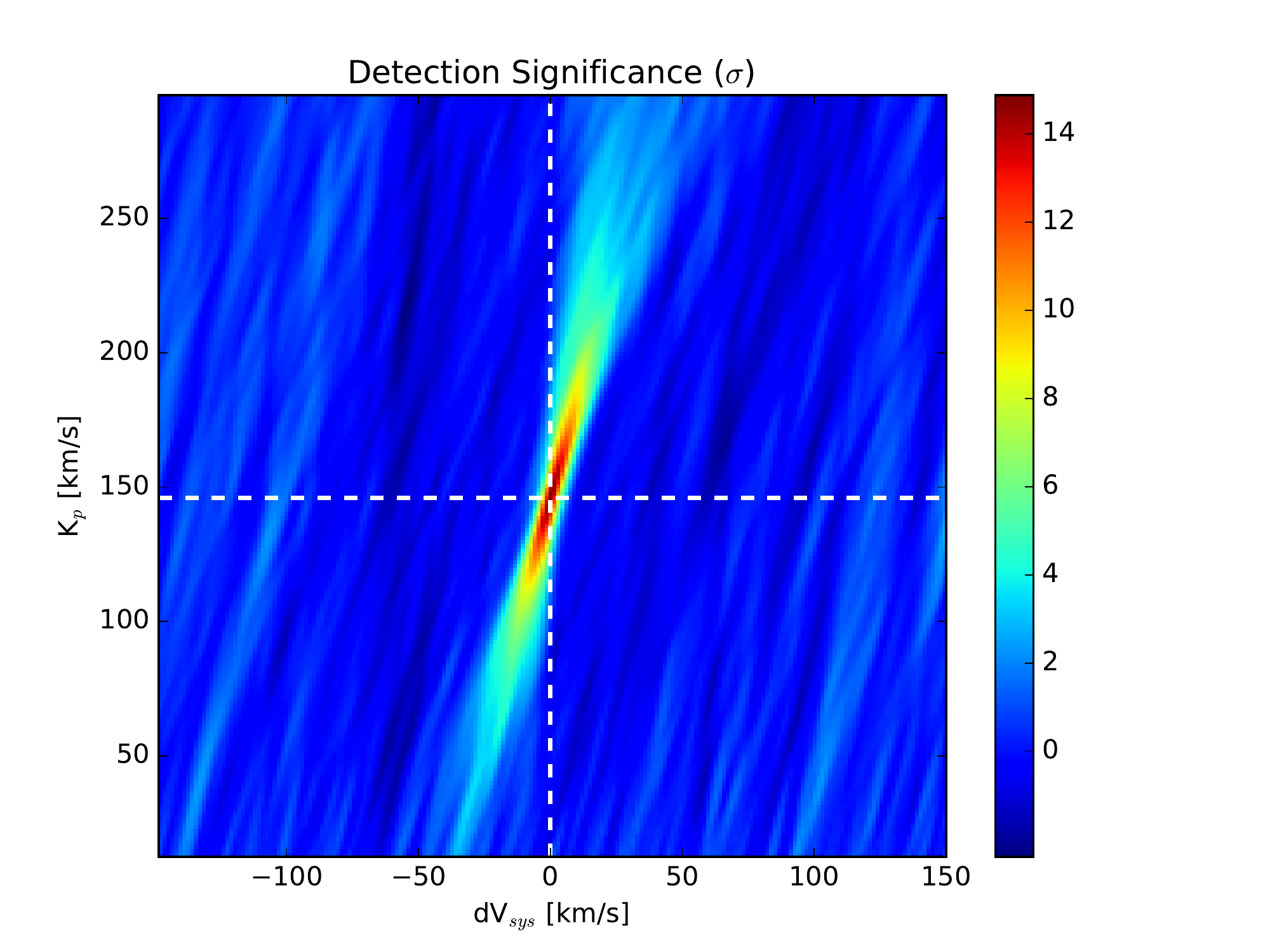}
    \caption{Planetary signal detected in the ``$K_p$-$dV_{sys}$'' plane.  The x-axis is the differential system velocity (0 km/s represents no deviation from the known time-dependent system velocity) and the y-axis is the maximum radial (Keplerian) velocity of the planet. The figure is generated by summing the cross-correlation coefficient at all phases for each combination of $K_p$ and $dV_{sys}$ and normalizing by the ``noise'' which is an average of the cross-correlation coefficients over a portion of velocity space far from the planetary signal.  This noise/planetary parameter setup results in a $\sim$12$\sigma$ detection of the planetary atmosphere at the ``known'' velocities (white dashed-lines). }
    \label{fig:demo_kpvsys}
\end{figure}

The standard approach at interpreting \hires\ observations would be to store the cross-correlation coefficients for each velocity, each spectrum, each detector and each night, and determine the ``detection significance'' of the planetary signal in the planetary-systemic velocity plane via the total cross-correlation coefficient summed over all observed orbital phases. This sum could be weighted to account for the different signal-to-noise of each spectrum or variable telluric or planet signal of CRIRES detectors, introducing some sort of subjectivity to the process. The significance of the detection is then measured either by taking the ratio between the peak value of the total cross-correlation and the standard deviation of the cross-correlation coefficients around it, or by applying a more sophisticated t-test on the distributions of cross-correlation coefficients around and far from the planet radial velocity. If a high significance (usually larger than $\sim 4\sigma$) is obtained at the expected velocity pair, then the planetary atmosphere is said to be detected \citep{sne10,bro12,dek13,bir13,bir17,schwarz15}. Figure \ref{fig:demo_kpvsys} shows the detection significance for the simulated planet signal (with the underlying CO and H$_2$O abundances given in Table \ref{tab:model_params}) which should aid in the qualitative mapping between standard methods and the constraints obtained through our retrieval method. Given our simulated planetary/stellar/noise properties, we would obtain a $\sim$14$\sigma$ detection of the planetary signal in the planetary-systemic velocity plane at the input velocities. This fairly optimistic signal compared to the majority of past CRIRES detections is a combination of a model particularly rich in spectral lines, of the fact that we have neglected detector readout and thermal noise, additional photon noise from sky emission, and effects of damaged pixels and regions on the detector.

In this new framework, to obtain the total signal from the planet we just need to co-add all the log-likelihood values as function of time, detector, and/or night of observation. Contrarily to previous studies, there is no need to weight cross-correlation functions anymore, because our likelihood contains the data and model variances ($s_f^2$ and $s_g^2$, respectively), hence it intrinsically incorporates the variable SNR of the observations. This is another significant advancement of our framework and it adds objectivity to the retrieval process.
In the following sections we perform a series of exploratory experiments with our novel \hires\ retrieval approach on this processed simulated data set.


\subsection{A Simple \hires\ ``Free'' Retrieval Example}\label{sec:retrieval}
\begin{figure*}
\centering
\includegraphics[width=1.8\columnwidth]{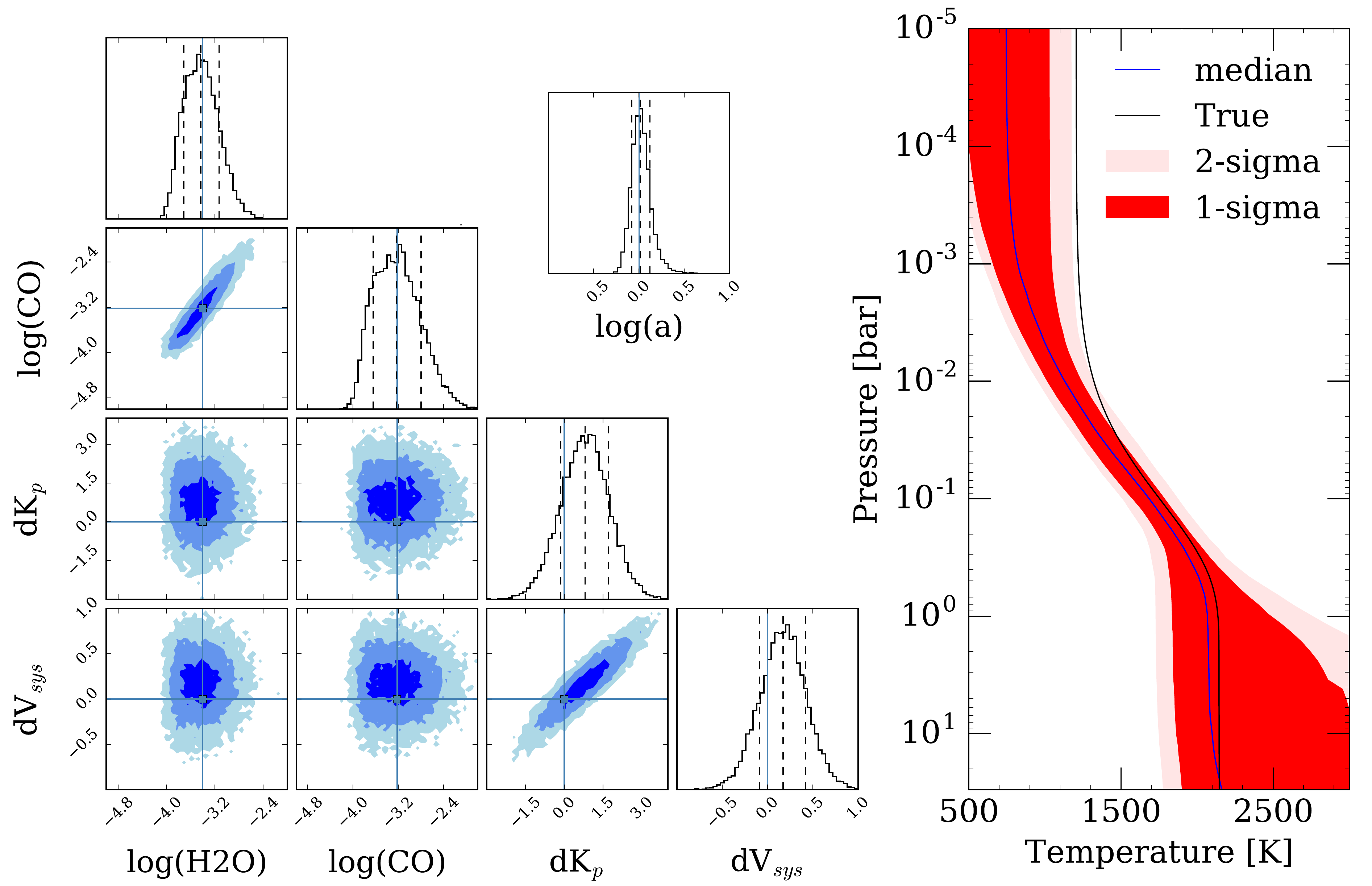}
\caption{Retrieved constraints from the simulated HDS data set.  Constraints are summarized with a corner diagram for the abundances and velocities and the \tp\ profile spread reconstructed from the parametrization, as is common in the \lores\ retrieval literature \citep[e.g.][]{lin13a}.  We also show the marginalized constraint on the retrieved scale factor, $a$, to show that we indeed recover the expected maximum likelihood estimator ($\hat{a}=1$ or $\log\hat a=0$).  The water abundance is constrained to $\pm$0.3 dex, CO to $\pm$0.43 dex, and the velocities to $\pm$0.93 and $\pm$0.26 \kms\ respectively. The difference between log(CO) and log(\water), a proxy for the C-to-O ratio (not shown) is constrained to $\pm$0.18 dex.  The input values, or ``truths'' are indicated with the vertical light blue lines and box in the corner plot. All corner plots in the remainder of the manuscript were generated with a modified version of the {\tt corner.py} routine}.
    \label{fig:demo_retrieval}
\end{figure*}

Figure \ref{fig:demo_retrieval} summarizes the retrieved properties on the simulated \hires\ data set (under the {\it free} retrieval assumption parameterized with the variables given in Table \ref{tab:model_params}).  These constraints are quite remarkable--on the order of 0.5 dex for CO and \water, despite the inclusion of realistic noise sources and common atmospheric retrieval parameterizations utilized in \lores\ data interpretations. It is encouraging to note that there is no bias in the retrieved parameters, including the scale factor $a$\footnote{Leaving $a$ as free parameter allows us to verify that $a=1$ is an unbiased choice.}, beyond what is expected due to the random noise-instantiation ($N_\mathrm{mod}(\lambda,t)$ in equation \ref{eq:noisy_model}).  

We notice some curious, but expected degeneracies. Firstly, the water and CO abundances are strongly correlated. Increasing the water abundance would require an increase in the CO abundance to maintain an acceptable log-likelihood.  The reason for this degeneracy is that the retrieval tries to preserve the ratio between the CO and the water lines.  Increasing both together, over some range, preserves this ratio.  This suggests that the \hires\ data is highly sensitive to the abundance {\sl ratios}.  In fact, the precision on $\log$(CO)-$\log$(\water), a good proxy for C/O, is $\pm$0.18 dex, about a factor of two better than for absolute abundances. Another noteworthy, but unsurprising degeneracy is between the two velocities. This is simply a reproduction of the degeneracy between \kp\ and \vsys\ seen in the total cross-correlation signal/detection significance (Figure \ref{fig:demo_kpvsys}) that is easily lifted by repeating observations at different phase range. At least in this example, there do not appear to be any degeneracies between the abundances and the velocities.  Additional degeneracies appear amongst the 3 parameters describing the \tp\ profile (not shown) that reflect the retrievals desire to maintain the temperature gradient over the 1 bar to 10 mbar region of the atmosphere.     

It is worth noting that the tightest constraint on the water abundance through HST WFC3 emission spectroscopy (1.1 - 1.7 $\mu$m) is $\pm$0.6 dex \citep[WASP-43\,b,][] {Kreidberg2014}.  The high-resolution data, in particular this narrow slice of $K$-band spectra from 2.29 - 2.34 $\mu$m, shows potential to constrain not only the water abundance to a higher precision, but also the CO abundance unobtainable with WFC3 alone.  

The sensitivity to absolute abundances is perhaps the most unexpected outcome of this framework. One would expect that normalizing the data as described in Section~\ref{sec:analysis} removed any sensitivity to absolute fluxes, hence absolute abundances. However, this sensitivity is partially recovered by choosing $a=1$ in Section~\ref{sec:newlogL}. This still does not set the absolute continuum, however it sets the absolute planet line depth compared to the continuum of the star. Conditional to a proper normalization of the model (in planet/star units), this choice is completely consistent with the analysis of transit or eclipse spectroscopy at low spectral resolution. Furthermore, if significantly uncertain, the normalization parameters (stellar temperature, planet/star radii) can be inserted as free parameters with appropriate priors into the framework at nearly no computational cost.

We acknowledge that large uncertainties in the absolute scaling (or stretch) in the line-to-continuum contrast, due to either uncertain planet/star properties or inaccurate telluric removal would seemingly inhibit absolute abundance determinations.  However, simply scaling the line-to-continuum ratio is not the same as increasing or decreasing the absolute abundances.  Changing absolute abundances will not affect all lines equally due to their different relative positions on the curve-of-growth. Strong lines may saturate whereas weak lines will continue to increase their contrast relative to the continuum.  It is this relative line depth (and shape) behavior of lines of the same gas that permits the absolute abundance constraints. Certainly there will be regimes where this degeneracy is prohibitive, such as in nearly isothermal atmospheres or ultra-low abundances of all gases over a given pass-band. However, such scenarios are likely to be rare.

\begin{figure*}
	\includegraphics[width=2\columnwidth]{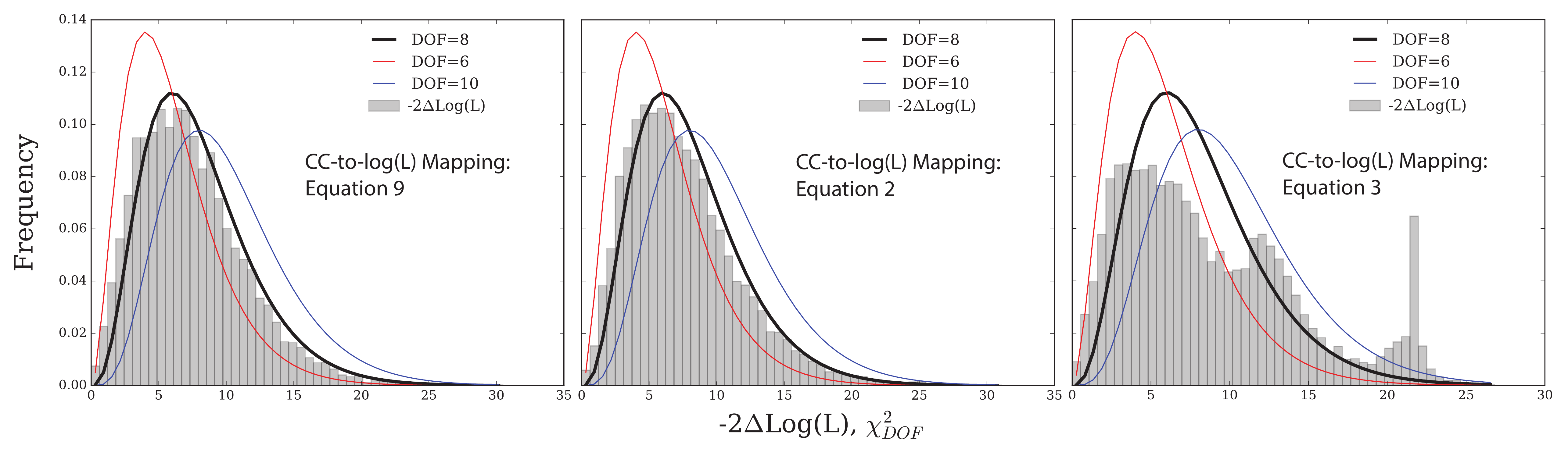}
    \caption{Test of the robustness of our \cctologl\ mapping. Wilks' theorem states that the test statistic $-2\Delta\log(L)$ for an $M$-parameter (in this case $M=8$) estimator should follow a chi-square distribution with $M$ degrees of freedom (DOF). Here, $\Delta$log($L$) is the difference in the log-likelihood between the maximum likelihood and all other likelihoods within the posterior probability distribution. Histograms {\bf of} this test statistic computed with our \cctologl\ mapping (Equation~\ref{eq:logL_CC2}, left panel) or with the mapping of \citet{zuc03} (Equation~\ref{eq:Zucker_Map}, middle panel) closely follow a $\chi^2_{8}$ distribution as expected. In contrast, if we incorrectly apply the mapping of \citet{loc14} to our data (Equation~\ref{eq:Lockwood_Map} and Section~\ref{sec:past_map}) the resulting histogram does not appear to follow any $\chi^2_\mathrm{DOF}$ distribution.} 
    \label{fig:wilks_test}
\end{figure*}

It is also important to assess the statistical validity of the cross-correlation to log-likelihood mapping. Wilks' Theorem \citep{wilks1938} suggests that the distribution of delta-log-likelihoods (specifically, $-2 \Delta \log(L)$ differenced from the maximum $\log(L)$) over a posterior probability distribution should follow a chi-square distribution with number of degrees of freedom equal to the number $M$ of parameters--in this example, $M=8$ free parameters (Table \ref{tab:model_params}). The left panel in Figure \ref{fig:wilks_test} shows the distribution of $-2 \Delta \log(L)$ drawn from the posterior probability distribution of our test case under the \cctologl\ mapping described in Section~\ref{sec:newlogL}.  The histogram of $-2 \Delta \log(L)$ correctly follows a $\chi^2$ distribution with 8 degrees of freedom suggesting an appropriate mapping.  

As an additional check, though not shown, we performed 100 independent \hires\ retrievals on 100 separate photon noise instances ($N_\mathrm{mod}(\lambda,t)$, but under the same telluric and planetary model-$F_\mathrm{mod}(\lambda,t) $). From this experiment we found that the distribution of parameter means (due to random noise scatter) agreed with what was expected from the uncertainties derived on an individual retrievals. In addition, the deviations from the truth values are random in the parameter space, i.e. they occur in random directions according to the particular noise instance. This fact strongly points to the absence of biases. The success of these robustness tests should not be surprising as the $\log(L)$ given by equation \ref{eq:logL_CC2} derives directly from inserting the relation between data and model in equation \ref{eq:logL_generic} and carrying out the algebraic passages without approximations.

\subsection{Comparison of \cctologl\ Mappings}\label{sec:logL_distr}
\begin{figure}
	\includegraphics[width=\columnwidth]{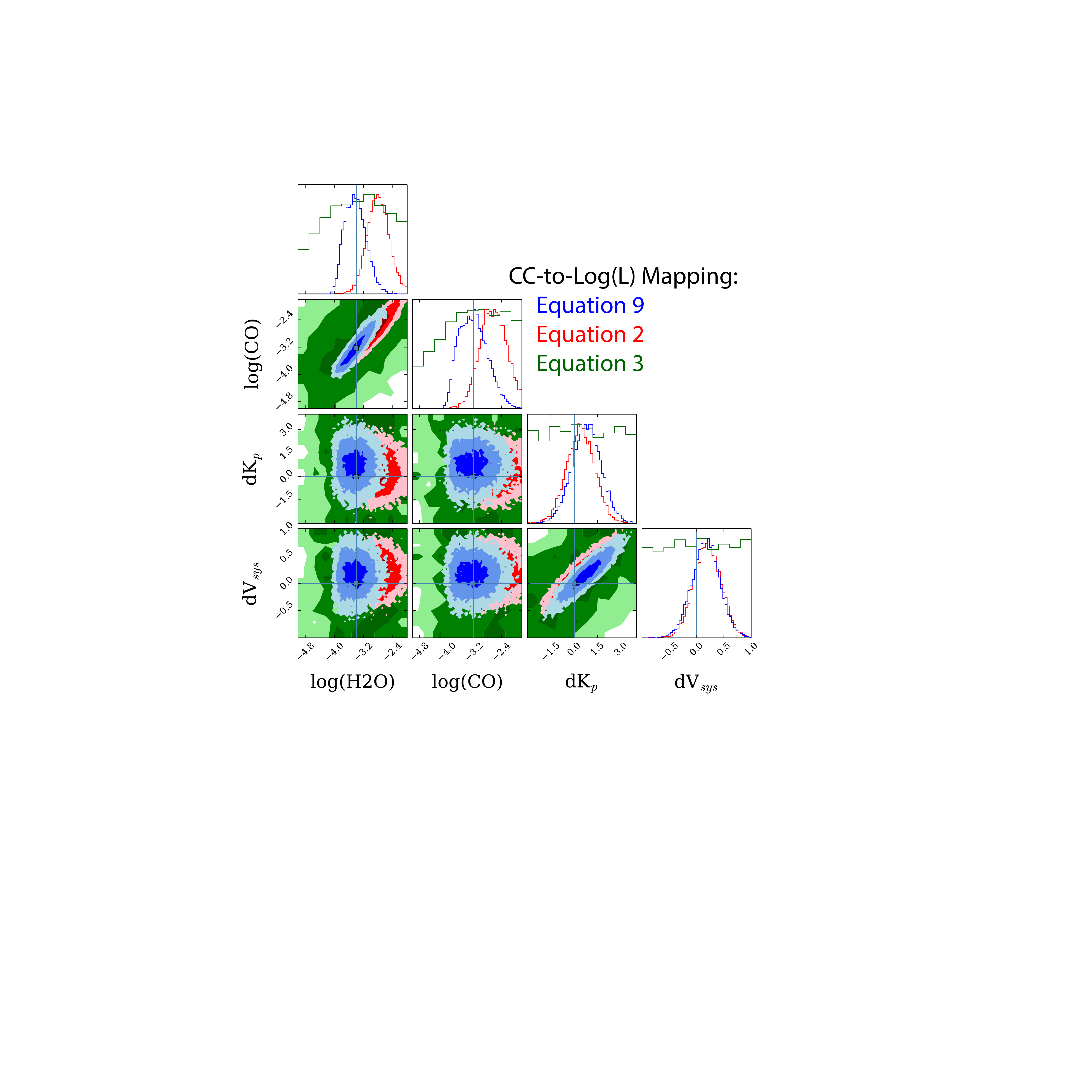}
    \caption{Comparison of constraints obtained under different \cctologl\ mappings on the same simulated dataset.  The mapping derived in this work (Eq.~\ref{eq:logL_CC2}) is shown in blue (same as Figure \ref{fig:demo_retrieval}), \citet{zuc03} in red, and \citet{loc14} in green.  Our mapping and \citet{zuc03} result in similar constraints, albeit with a parameter bias when using the \citet{zuc03}.  The \citet{loc14} mapping provides virtually no constraint on our dataset, which is expected since such formalism is not applicable to our CRIRES data (see Section~\ref{sec:past_map}). Note, the parameter prior ranges are broader than the plot axes.  }
    \label{fig:compare_logL}
\end{figure}

Figure \ref{fig:compare_logL} compares the constraints obtained under the three different mappings.  The mappings derived through equation~\ref{eq:logL_CC2} in this work and equation~\ref{eq:Zucker_Map} in \citet{zuc03} provide comparable parameter constraints.  Nevertheless the \citet{zuc03} mapping results in a bias in the medians of the retrieved H$_2$O and CO abundances.  These differences arise because the two mappings respond differently to a particular noise instantiation due to the inclusion (or lack-there-of) of the $s_{f}^2$ and $s_{g}^2$ terms.  As with our mapping, we find that the distribution of $-2\Delta\log(L)$ derived from the \citet{zuc03} mapping agrees well with a chi-square distribution of 8 degrees of freedom, as required by Wilks' theorem (Figure \ref{fig:wilks_test}, left and middle panels). As an example of the importance of using the correct \cctologl\ mapping, we also show the consequence of applying equation~\ref{eq:Lockwood_Map} {\sl incorrectly} to our data. In Section~\ref{sec:past_map} we explained that the mapping described in \citet{loc14} is only valid for stationary planet signals, and indeed forcing such mapping on our data analysis results in very broad, virtually non-existent constraints on all of the parameters (Figure~\ref{fig:compare_logL}, green contours). Furthermore, the resulting distribution of $-2\Delta\log(L)$ clearly does not follow a chi-square distribution (Figure \ref{fig:wilks_test}, right panel) confirming that this mapping is not appropriate for spectral time sequences where the planet radial velocity changes rapidly.  

\subsection{Impact of \water\ Cross-Sections}\label{sec:h2o_xsec}
\citet{bro17} explored the potential impact of incomplete or incorrect line lists on the cross-correlation signal. That work concluded that these uncertainties were not important mainly because water vapor was not detected in the high-resolution spectra. Having tested simulated spectra with both CO and \water\ in this work allows us to re-assess the importance of cross sections in a completely controlled environment, and we present in this Section the main results. 

Our simulated spectral sequence is the same as in Section \ref{sec:retrieval} and it uses \water\ cross sections calculated from \citet{Freedman2014} based upon the \citet{PS97} line-list. We run two separate retrievals on this spectral sequence. The former uses the same absorption cross sections and should therefore result in unbiased parameters. The latter uses absorption cross sections generated from HITEMP \citep{hitemp2010} via the HITRAN HAPI routine \citep{hapi} and is based on the \citet{Barber06} line-list. These cross sections are computed at 0.003 cm$^{-1}$ sampling resolution, 100 cm$^{-1}$ Voight wing cut-off. They assume ``air'' broadening over a well sampled pressure and temperature grid \citep{Freedman2014,Freedman2008}. While there are subtle differences\footnote{The difference in line widths due to air vs. H$_2$/He pressure broadening is well below the CRIRES instrumental resolution} in the line profile assumptions between the two sets of absorption cross-sections, the main impact on HRCCS is the variable line position arising from different choices of line-lists. We note that \citet{shab11} explored the differences in these two line lists at low resolution and found a negligible difference.   

\begin{figure}
	\includegraphics[width=\columnwidth]{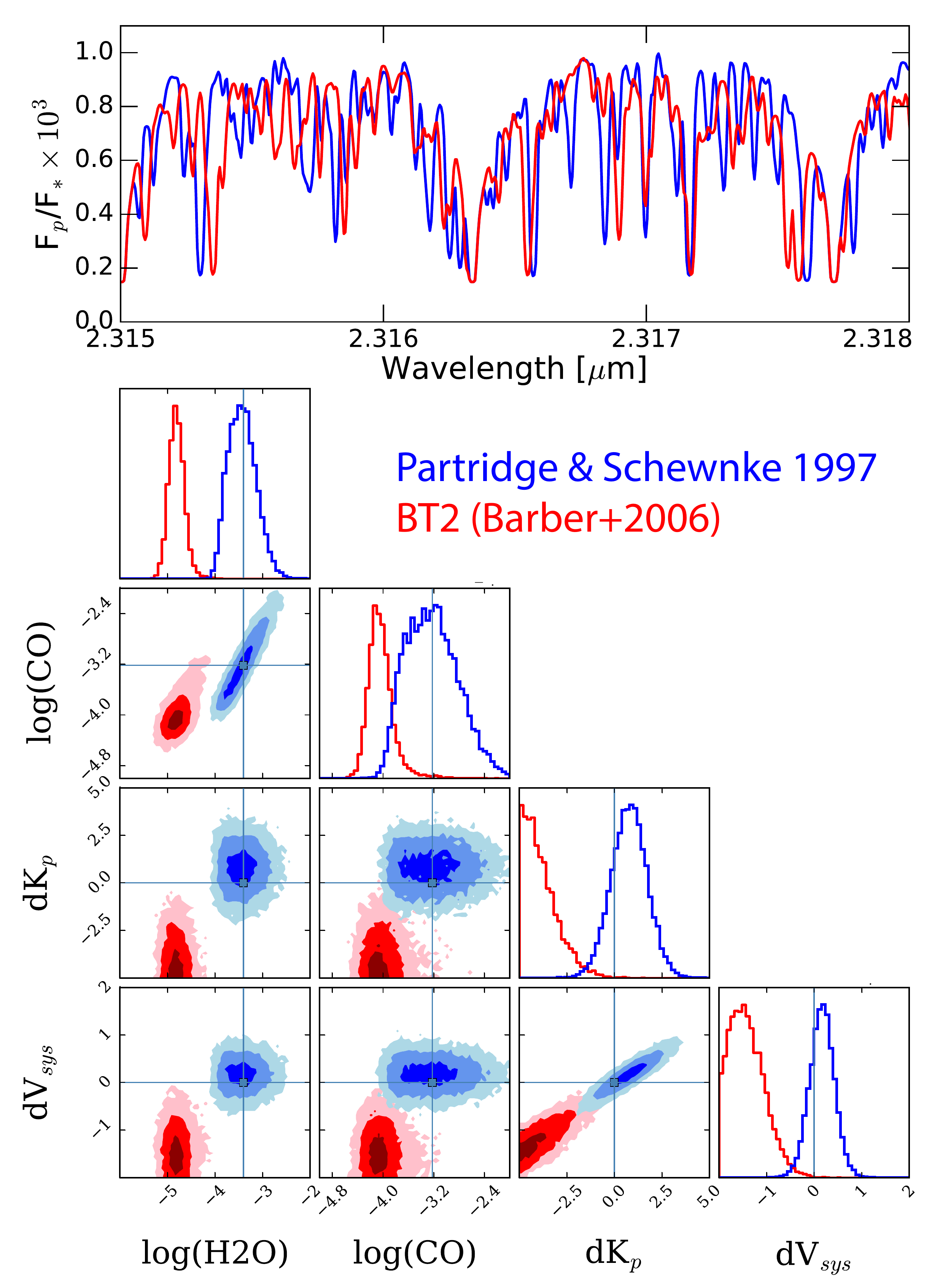}
    \caption{Impact of the line-list choice on the retrieved parameters.  We retrieve on the same synthetic dataset under two different line-list assumptions, in blue, \citet{PS97} implemented with in the \citet{Freedman2014} cross-sections (used to generate ``true" model) and HITEMP based upon the BT2 \citep{Barber06} line list (red). The top panel compares spectra (using the same inputs given in Table \ref{tab:model_params}) generated with the different absorption cross section databases over a small portion of the K-band.  The lines are clearly haphazardly shifted.  The posterior probability distribution summarized in blue is the same as in Figure \ref{fig:demo_retrieval}.  It is clear that there are significant parameter biases and uncertainty differences when retrieving with different sets of water absorption cross sections utilizing different underlying line lists.   }
    \label{fig:linelists}
\end{figure}

Figure \ref{fig:linelists} shows the impact of the line-list/cross-section assumptions.  By eye (top panel) it is easily seen that the two spectra do not perfectly overlay on each other at the resolutions attainable by CRIRES. The line position differences are not uniform in wavelength, suggesting that these differences cannot be compensated by a single velocity offset. These inconsistencies, when combined over the entire CRIRES $K$-band, result in substantial biases in the retrieved parameter distributions, which are offset by many sigma from their true state (Figure~\ref{fig:linelists}). The constraints on the abundances are also much tighter when using the ``incorrect'' line-list.  It is a reasonable question, in the case of actual data, to ask ``which line list is correct?''.  Unlike LRS data, with HRCCS data it is difficult to obtain a ``visual'' model fit to make such assessments. We therefore rely upon the Bayesian evidence to guide us (a natural output of the nested-sampling algorithm used in this work).  The log-Bayes factor between the model using the ``correct'' line-list  (blue in Figure~\ref{fig:linelists}) and the model using the ``incorrect" line-list (red in Figure~\ref{fig:linelists}) is 147.4. Note that there is no change in the number of parameters between the two models.  Since a log-Bayes factor larger than 5.0 on the Jefferey's scale is considered significant, this extreme difference overwhelmingly favors the model utilizing the ``correct" line-list.  This conclusion would not be apparent from the posterior alone, hence it is always important to compare the model evidences. 

Performing the same experiment within the chemically-consistent model framework results in a similar degree of bias (in metallicity and carbon-to-oxygen ratio) and large Bayesian evidence differences. This suggests line list biases may exist regardless of the model parameterization.

It is important to note that if our simulated dataset was calculated with cross sections from \citet{hitemp2010}, the results would have been opposite, i.e. the retrieval with \citet{Freedman2014} cross-sections would have been biased. This is an important point to make, because although there is no extensive testing in the literature regarding the choice of line lists at high spectral resolution, all the past \water\ detections with CRIRES were achieved with the line lists from \citet{hitemp2010}. When CO and \water\ were detected simultaneously, measured radial velocities (\kp\ and \vsys) were consistent between species, which suggests unbiased water determinations with this database. The only explicit test of different line lists in the literature, besides that of \citet{bro17}, is mentioned in \citet{flowers2018}. In the latter paper, water opacities from \citet{lupu14} (which are similar to those in \citet{Freedman2014}) did not produce any correlation signal with CRIRES transmission spectra of HD~189\,733~b, in contrast to the $>5\sigma$ detection of \citet{bro16} obtained with the HITEMP database.

The results of this Section are rather alarming and we should take this as a further warning that {\sl all} \hires\ interpretations are going to be strongly dependent on the choice of line-lists used in the model templates. From a purist retrieval modeling perspective,  line list properties (positions, broadening, strengths) should be parameterized so as to appropriately marginalize over them within an HRCCS retrieval. However, this is extraordinarily unwieldy as this would slow down a forward model computation to the point of being unusable within a retrieval framework. Furthermore, a proper way of readily parameterizing these effects other than the standard brute-force line-by-line computation does not yet exist.  Such an approach, to be feasible, would likely have to make assumptions and approximations that would introduce additional uncertainties to the point of obviating its purpose.  Instead, we {\em strongly} advocate for further laboratory, astrophysical, and {\it ab initio} efforts to determine accurate line positions, intensities, and broadening for exo-atmosphere relevant molecules and conditions \citep[e.g.][]{Fortney16}. 
We would anticipate that as we push towards cooler temperatures or more ``€œearth-like''€ conditions, we will have the opportunity to better quantify and reduce line-lists uncertainties by validating the output of radiative transfer calculations on spectra of our own planets or other solar system planets. More accurate line lists will likely increase the overall level of correlation with terrestrial exoplanet spectra and minimise biases in retrieved parameters, which will be crucial for the robust assessment of habitability and bio-signatures.

\subsection{Combined \lores\ and \hires}\label{sec:lds_hds_combi}
Once a statistically robust \cctologl\ mapping is achieved, one can trivially combine information from various datasets. One such potentially useful combination is that between low resolution spectroscopy with HST WFC3 and high resolution spectroscopy with VLT CRIRES or Keck NIRSPEC.  In this Section, we combine simulated HST WFC3 data (loosely based upon the typical emission observations--30ppm/channel at 0.035 $\mu$m bins) with the above CRIRES $K$-band data within the chemically-consistent modeling framework (whereby metallicity and C/O are retrieved under the assumption of equilibrium chemistry).

Trivially, to combine the inference from different sets of data we just sum their log-likelihoods:
\begin{eqnarray}\label{eq:combined_logL}
\log(L_{tot}) = \log(L_\mathrm{HDS}) + \log(L_\mathrm{LDS}),
\end{eqnarray}
where $\log(L_\mathrm{HDS})$ is given by Eq. \ref{eq:logL_CC2} and 
\begin{eqnarray}
\log(L_\mathrm{LDS}) = -\frac{1}{2}\chi^2
\end{eqnarray}

\begin{figure}
	\includegraphics[width=\columnwidth]{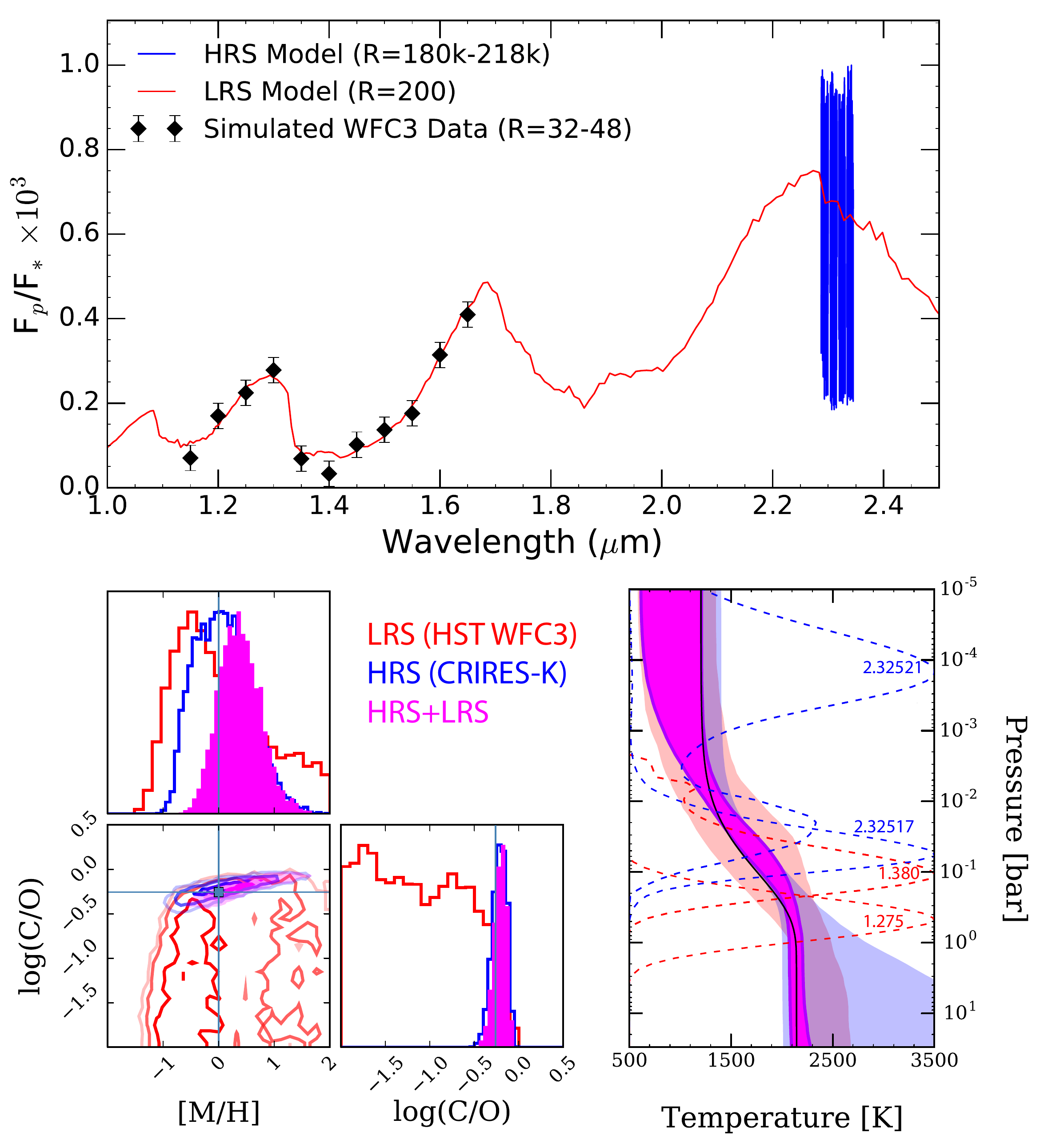}
    \caption{Simulated combined observations with HST/WFC3 (black dots, top panel) and VLT/CRIRES around 2.3$\mu$m (blue lines, top panel). The bottom panels show the posterior distributions for planet metallicity ([M/H]), carbon-to-oxygen ratio (C/O), and \tp~profile (bottom-right panel, summarized with the 68\% confidence intervals and obtained by running our framework on the HST data alone (red curves), VLT data alone (blue curves) and on the combined dataset (magenta curves).  The dashed curves (blue=CRIRES, red=WFC3) are the temperature Jacobians at the indicated wavelengths (on and off band/line). In general HST WFC3 probes a relatively deep and narrow region. In contrast, the high dynamic range in the CRIRES spectrum permits broad altitude coverage. Combining low- and high-resolution spectra leads to a substantial improvement in the precision of these measurements.  }
    \label{fig:LDSHDS_retrieval}
\end{figure}

The top panel of Figure~\ref{fig:LDSHDS_retrieval} shows the spectral regions covered by these simulated data, and is representative of observations that would have been already feasible 5 years ago.
The bottom panels show the retrieved planet properties by running our framework on the two datasets separately (WFC3 only in red, CRIRES K-band in blue, and on their combination in magenta derived through the combined log-likelihood function in Eq. \ref{eq:combined_logL}) under the {\it chemically-consistent} model parameterization (Table \ref{tab:model_params}). 

The simulated HST WFC3 data alone are able to constrain the planet metallicity ([M/H]) to within 1.7 dex (68\% confidence), slightly better than expectations from published results \citep{lin16}.  This uncertainty is primarily driven by the degeneracy between the C/O and metallicity in producing the same water abundance (\water\ is the main measurable gas over the WFC3 pass-band). In contrast, the K-band CRIRES observations alone result in a 1.0 dex (68\% confidence) constraint, nearly a factor of 2 better than obtained with WFC3.  However, the power in these particular HRCCS observations is in their ability to constrain the carbon-to-oxygen ratio--to within 0.2 dex (68\% confidence), driven by the enhanced sensitivity to CO-to-H$_2$O line ratios at hi-resolution.  WFC3 observations alone naturally struggle to constrain the C/O, upper limits only, due to the lack of presence of carbon bearing species in this wavelength range.  In this particular setup, however, combining these two datasets through a joint likelihood does little to improve beyond the HRCCS constraints alone. In fact, the abundant parameter space mass present at higher metallicities in the WFC3 only scenario, tends to pull the joint constraint towards these higher metallicities.  The HST observations, however, help in constraining the TP profile as they are weighted towards deeper layers of the atmosphere.  Certainly this only a single example of the combination of these two datasets; perhaps unfairly at the detriment to the HST observations.  We imagine a perhaps more synergistic setup might be with LRS mid-infrared ($>5\mu m$) of which are best obtained from space (e.g., JWST/MIRI), combined with ground based near IR HRCCS observations.  This would provide maximum leverage of the strengths of each approach.


Since we are testing a dataset based on a planet without a thermal inversion, in terms of \tp~profile the two dataset probe reasonably similar regions of the planet photosphere (Figure~\ref{fig:demo_retrieval}, bottom-right panel), with only the CO line cores absorbing at significantly lower pressures (up to 10$^{-4}$ bar). Nevertheless, the combined dataset allows us to precisely determine both the atmospheric lapse rate and the photospheric temperatures\footnote{We are aware that these may be overly optimistic constraints due to the particular choice of \tp-profile parameterization. Certainly the temperature is not constrained this precise over the entire atmospheric column shown.}. In the case of high-altitude thermal inversion layers (temperature increasing with altitude), we anticipate that the complementarity between low- and high-resolution spectra would be even more evident, with the two constraining mostly the lower and the upper atmosphere, respectively.

\subsection{Influence of Missing Gases}\label{sec:missing_gas}
As with LRS retrieval analyses, failure to include all of the key relevant opacity sources in an HRCCS model could result in significant biases in inferred atmospheric properties.  We perform a simple experiment to explore the impact that unaccounted gases can have on retrieved atmospheric properties. 

\begin{figure}
	\includegraphics[width=\columnwidth]{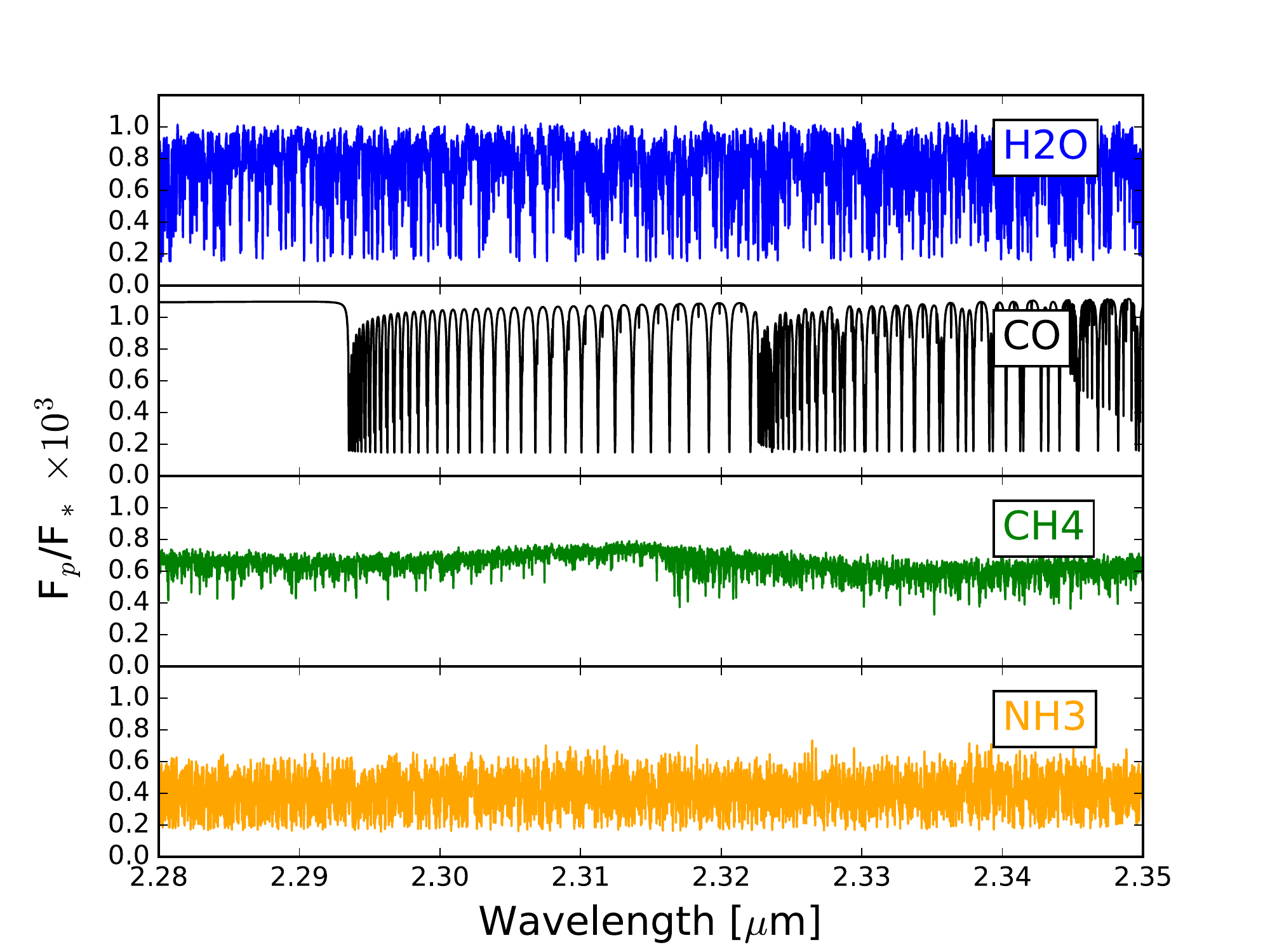}
    \caption{Spectral components of the important absorbers over the CRIRES K-band.  H$_2$O and CO have strongest influence, whereas CH$_4$ and NH$_3$ contribute to mostly continuum absorption due to their reduced line-to-continuum contrast. These spectra are at the native line-by-line cross-section resolution.   }
    \label{fig:gas_spectra}
\end{figure}

We utilize the same set up as in Section \ref{sec:retrieval} (the ``free retrieval'') but include in the simulated ``true'' spectrum additional molecular opacities due to CH$_4$, NH$_3$, CO$_2$,  and HCN (log(Mixing Ratio)=-5.0, -5.0, -4.5, -9, respectively), though the latter two gases have little impact. Figure \ref{fig:gas_spectra} shows the abundance weighted contributions of the dominant gases to the spectrum.  

\begin{figure}
	\includegraphics[width=\columnwidth]{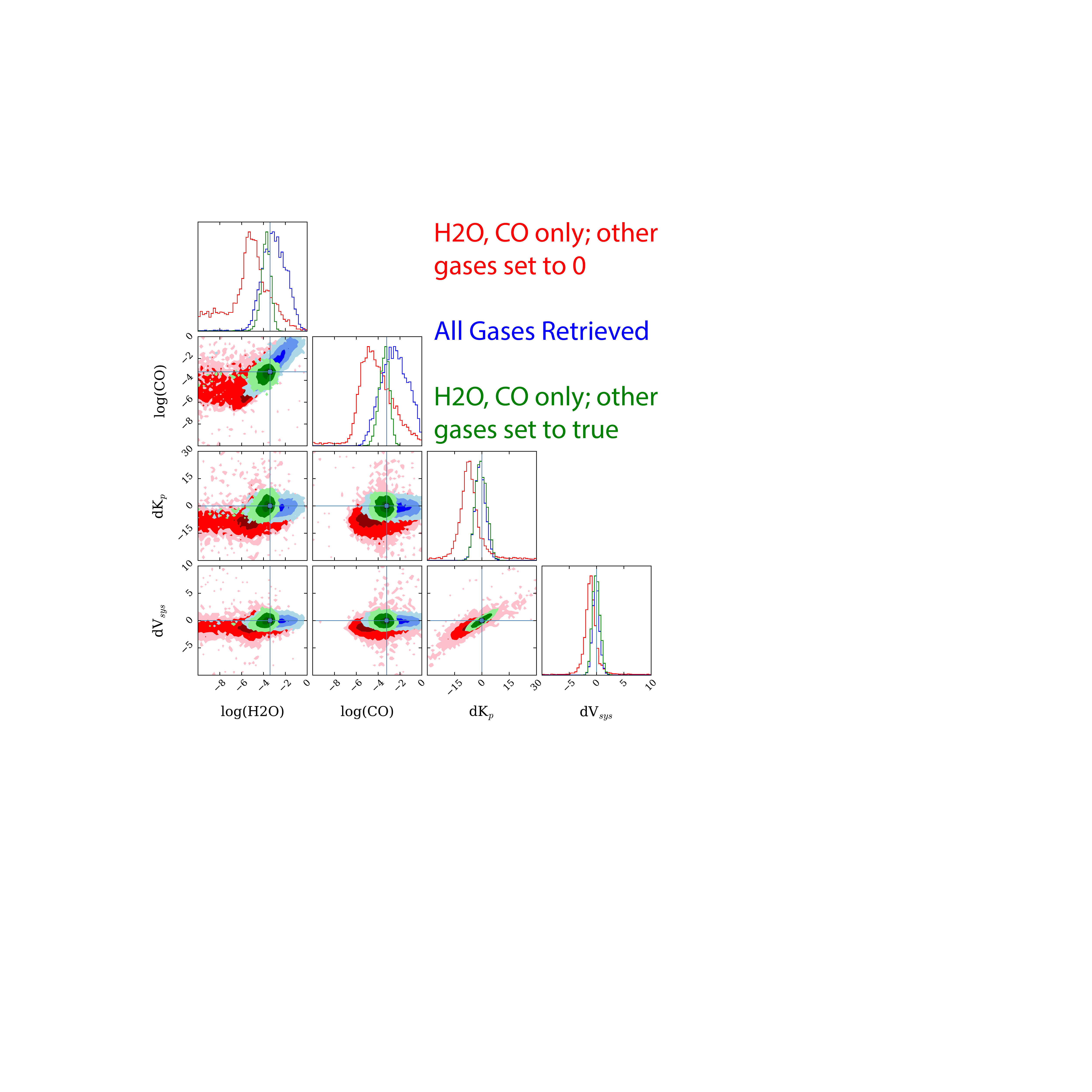}
    \caption{Influence of unaccounted for absorbers on the water, CO, and velocity constraints.  The underlying true model includes opacities from H$_2$O, CO, CH$_4$, CO$_2$, NH$_3$, and HCN in near solar proportions, though only H$_2$O, CO, CH$_4$, and NH$_3$ noticeably contribute.  The posterior distributions summarized in red represent the scenario for which only H2O and CO are retrieved and the other opacities are turned off.  This results in both large uncertainties and biases in the constraints.  The blue represents a full marginalization over all 6 gases.  The uncertainties are larger, but there is no bias. Finally, green represents a scenario where the other gases are fixed to their true input values but only H2O and CO are retrieved. This proudces the most precise constraints.  }
    \label{fig:missing_gas}
\end{figure}

We then perform 3 different retrievals (summarized in Figure \ref{fig:missing_gas}). The first scenario (red posterior distribution in Figure \ref{fig:missing_gas} ) zeros out the abundances of the other gases within the retrieval and fits for the standard ``free retrieval'' parameters as in Section \ref{sec:retrieval}. In other words, this scenario fails to account for all of the opacity sources.  The resulting constraints are biased and produce large uncertainties.  The water abundance is poorly constrained.  The second scenario (blue) retrieves for all 6 gases. That is, the wide uncertainties reflect the full marginalization over all of the included gases, but there is no bias.  This is the ``most correct'' of the 3 scenarios.  Finally, the last scenario (green) retrieves again the ``default'' set of parameters, but we assume we have perfect a-priori knowledge of the other 4 gases of which are fixed to their true input values.  Of course, this would rarely be the case. This results in artificially tight constraints, unsurprisingly as there are fewer ``free" gases to confuse water and CO.  The Bayesian evidence overwhelmingly favors, obviously, the latter scenario (log-Bayes factor of 22.8 between scenario 3 and 1, and 34.3 between 3 and 2). However, perhaps unfortunately, the first scenario (\water, CO only; other gases set to 0) is favored (11.5) over the second (all gases retrieved).  Relying upon the Bayes factor alone, in this situation, would lead one to conclude that the \water/CO only model is the correct one, when in fact it is not.  This is likely because of the unnecessarily large prior volume due to the inclusion of CO$_2$ and HCN in the retrieval despite their negligible impact on the spectrum (due to their low abundances).  In practice, as is now routinely done in LRS modeling analysis \citep[e.g.][]{kreidberg2015}, nested model comparison with sequential removal of gases relative to some parent model should be performed to determine the ``optimal" number of gases to include.

In summary, it is extremely important to be cognizant of all of the potential sources of opacity present over a given hi-res band-pass and how their lack of inclusion could bias atmospheric inferences.  This will indeed become increasingly important with broader spectral range instruments slated to come online in the not too distant future.


\section{Application to CRIRES HD~209458\,\lowercase{b} \& HD~189733\,\lowercase{b} K-band Data}
\label{sec:real_data}

\subsection{Reduction Process and Setup}
We applied the framework described in Section \ref{sec:framework} and Section \ref{sec:testing} to real K-band dayside spectra of exoplanets obtained with CRIRES at the VLT. We re-analyze the half night of observations of HD~189\,733 b published in \citet{dek13} and the two half nights of \pname\ presented in \citet{schwarz15} and \citet{bro17}, to which we point the reader for more information. Here we recall that these datasets were observed around the strong 2-0 ro-vibrational band of carbon monoxide at 2.3 $\mu$m, but also that they should contain additional opacity from water vapor. Although \water\ was indeed detected in the same CRIRES range for other exoplanets observed in dayside \citep{bro13, bro14}, it was not detected for these two planets. In the case of \citet{schwarz15}, only a tentative detection of CO was presented, whereas a more advanced weighting of the CRIRES detectors allowed \citet{bro17} to recover the signal of CO at a S/N = 5, but again no water. Recently, \citet{hawker18} have revisited the dataset with a different de-trending algorithm for telluric lines, and confirmed the detections of CO and \water.

Since we have substantial literature to support the potential of these data, in this work we apply the most objective analysis process by matching Steps 4-7 described in Section \ref{sec:analysis} and visualised in Figure~\ref{fig:demo_flow}. Contrarily to past work, this analysis does not require optimisation of de-trending parameters, and is therefore ideal to apply our framework to data as uniformly as possible. 

There is only one extra step required in the analysis of HD~189\,733 b data. With a spectral type of K1V, the parent star shows strong CO absorption lines in the CRIRES spectral range. These lines are not completely stationary in wavelength in the observer reference frame, because the barycentric velocity changes by about 0.5 \kms\ during the 5 hours of observations. This is due to the changing orbital and rotational radial velocity of the Earth compared to the center of mass of the solar system. Fortunately, we have devoted abundant work in the past to the correction of stellar CO lines from HD~189\,733. In this context, we apply the state-of-the-art three-dimensional modelling of the stellar photosphere described in \citet{magic2013} and \citet{chiavassa2018}, and implemented as in \citet{flowers2018}. We divide out this modeled stellar spectrum between Step 4 and 5 of the analysis (see Figure~\ref{fig:demo_flow}), i.e. just before the removal of telluric lines. We note that this modeling is not parametric. Being completely self-consistent, our stellar 3D models only assume an initial metallicity for the star, which is well constrained in the literature and has been also verified by inspecting the shape and depth of stellar CO lines in the CRIRES data. Therefore, there is no subjectivity in the correction of the stellar spectrum, as no extra fitting or optimization is required at this stage. 

One final caveat for the analysis of real spectra is that we do not have an accurate model for the temporal variations of the telluric spectrum, which is needed to replicate the stretching and scaling of planetary signals due to telluric removal. We thus store the fitted telluric absorption spectrum obtained through Steps 4-6 of the analysis and use it to process each of the tested models, following the same prescriptions as for the simulated dataset (Section \ref{sec:analysis}).

The forward model used for both objects deviates slightly from that used in the simulated case (Section \ref{sec:retrieval}) in that we add an optically thick gray cloud parameterized with a cloud-top-pressure (CTP, log($P_c$)) and explore an additional ``simple'' \tp~profile parameterization similar to that described in \citet{lin16}.  We also adjusted the prior upper bound on the irradiation temperature ($T_\mathrm{irr}$ in Table \ref{tab:model_params}) for each object to prevent un-physically hot temperatures. 

\subsection{HD-189733b Retrieval Results}
\begin{figure}
	\includegraphics[width=1\columnwidth]{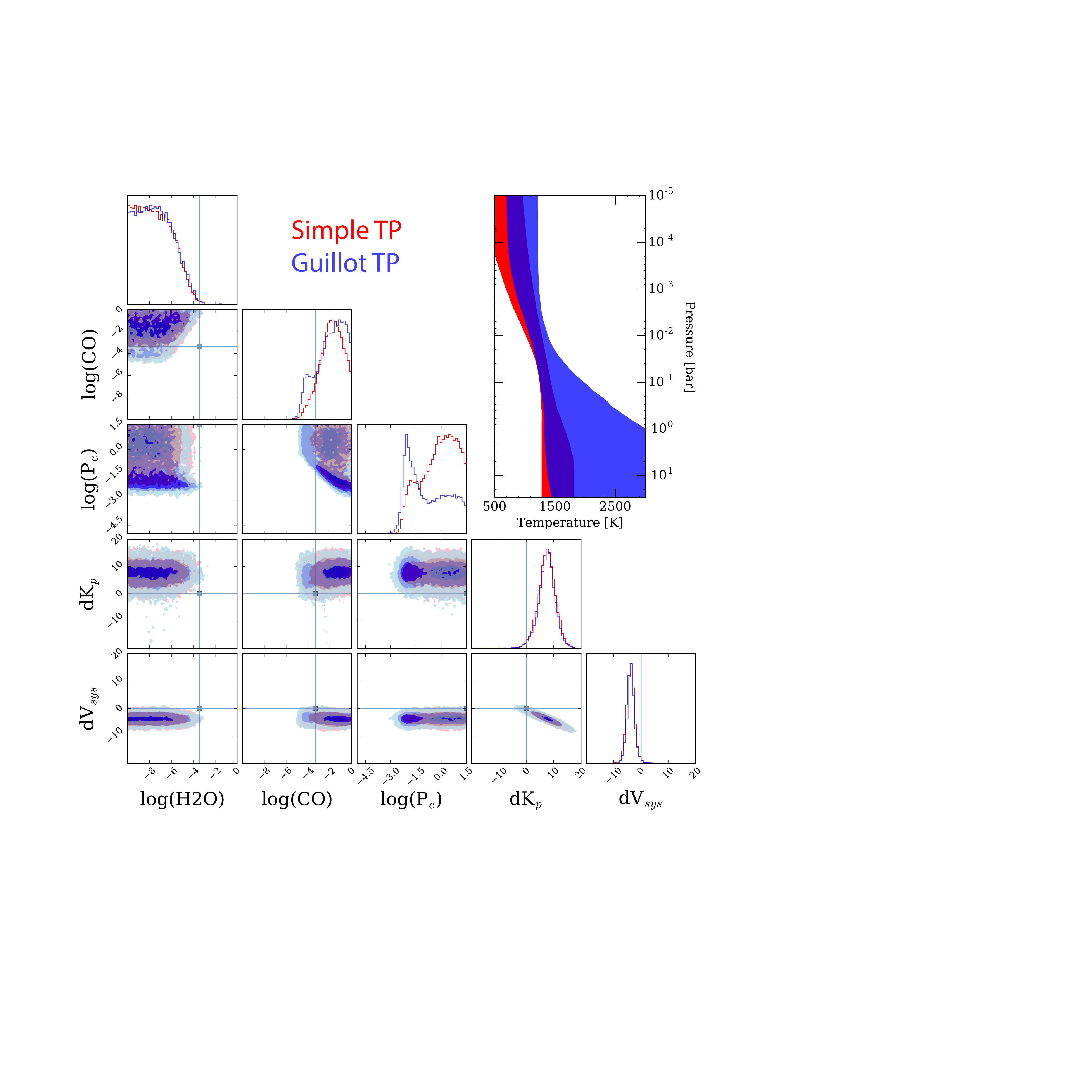}
    \caption{Summary of the HRCCS retrieval results for the CRIRES K-band dayside emission spectrum of HD 189\,733 b.  The blue histograms summarize the posterior under the ``default'' \citet{Guillot2010} TP-profile parameterization, and red the ``simple'' TP-profile parameterization described in \citep{lin16}.  The light blue lines/box within the 1D/2D histograms indicate the approximate mixing ratios predicted by thermo-chemical equilibrium at solar abundance and the 0-offset velocities. Only an upper bound on water, and a lower bound on CO are retrieved with the cloud and TP-profile parameterization having a negligible impact.}
    \label{fig:HD189_Summary}
\end{figure}

Figure \ref{fig:HD189_Summary} summarizes the HRCCS retrieval results within our framework.  We find only an upper limit on the water abundance, significantly lower than expected for solar elemental ratios, and a lower limit on the CO abundance but consistent with solar expectations.  The lower pressure limit on the CTP is consistent with a cloud-free day-side; however there is a notable degeneracy with the CO abundance whereby decreasing the CTP (higher altitude) results in an increase in CO abundance.  This is simply understood as the competition between the muting of the line-to-continuum ratio by the cloud and increasing line-to-continuum ratio with increasing CO abundance.  However,  not including the cloud has little impact on the CO abundance.  We also find that the abundances are largely insensitive to our choice of \tp~profile parameterization, even if the retrieved \tp~profiles themselves are different.  

The planet velocities are shifted by a few \kms\ from their nominal values.  We emphasize that it is important to marginalize over the velocities as uncertainties in orbital properties, especially years after their publication, can result in artificial velocity shifts. In fact even for well-known exoplanets such as HD 189\,733~b, errors in the quantities defined in equation~\ref{eq:pl_rv}, in particular \kp\ and $\varphi$, can lead to radial-velocity uncertainties of a few \kms, well above the sensitivity of these observations. A more subtle aspect is that many of the fundamental orbital parameters such as semi-major axis and orbital period can be correlated, hence using the reported error bars in the literature might lead to lower limits only on the final uncertainties.

In order to assess the uncertainties in planet velocities we compute the error bars on \kp\ from the stellar radial velocity amplitude and planet/star mass ratio, and the error bars on the orbital phase $\varphi$ from the time of mid-transit and the orbital period, all in the hypothesis of circular orbits. In the case of HD 189\,733 b, we obtain negligible errors in the orbital phase, but a $\Delta$\kp\ of about 6 \kms, which is sufficient to bring the retrieved \kp\ within the 1-$\sigma$ uncertainty. 

The obvious finding from the retrieval of HD 189\,733 b data is the strong detection of CO (lower abundance limit--4.67$\sigma$ according to the nested-sampling derived Bayesian evidence ratio) and non-detection of water (upper abundance limit). While this is surprising from solar elemental ratios and thermochemical arguments, it is not in the context of previous analysis of this dataset. Using a completely independent analysis on the same 2.3$\micron$ data, \citet{dek13} report a detection of carbon monoxide at 5$\sigma$ and no detection of water vapor. In contrast, the $L$-band data presented in \citet{bir13} show a clear detection of water vapor at 4.8$\sigma$. The spectra are taken with the same instrument (CRIRES), but at wavelengths with radically different water opacity. It is possible that a moderately low abundance of water vapor (e.g., VMR $\sim$ $10^{-5}$) produces a water spectrum too weak to be detected at 2.3 $\micron$ but sufficiently strong to dominate the spectrum at 3.2 $\micron$. Through transmission spectroscopy, water vapor is detected both at 2.3 $\micron$ \citep{bro16} and over the entire NIR \citep{bro18}, though the $abundance$ of \water\ is unreported in those works. We note that a recent independent re-analysis of the CRIRES spectra above \citep{cabot19} confirms the detection of CO at 2.3 $\micron$ and \water\ at 3.2 $\micron$. Additionally, HCN absorption is found in the 3.2-$\micron$ data.

Our retrieved \tp\ profile is qualitatively consistent with both \citet{dek13} and \citet{bir13} who both rule out a strong inversion due to the poor correlation with atmospheric models containing emission lines. 

Finally, as motivated by our findings in Section \ref{sec:h2o_xsec}, we again explored the impact of the line-list choice.  Unsurprisingly, because there is a lack of detection of water, we found virtually no difference in the posterior probability distribution.

\subsection{HD-209458b Retrieval Results}
\begin{figure}
	\includegraphics[width=1\columnwidth]{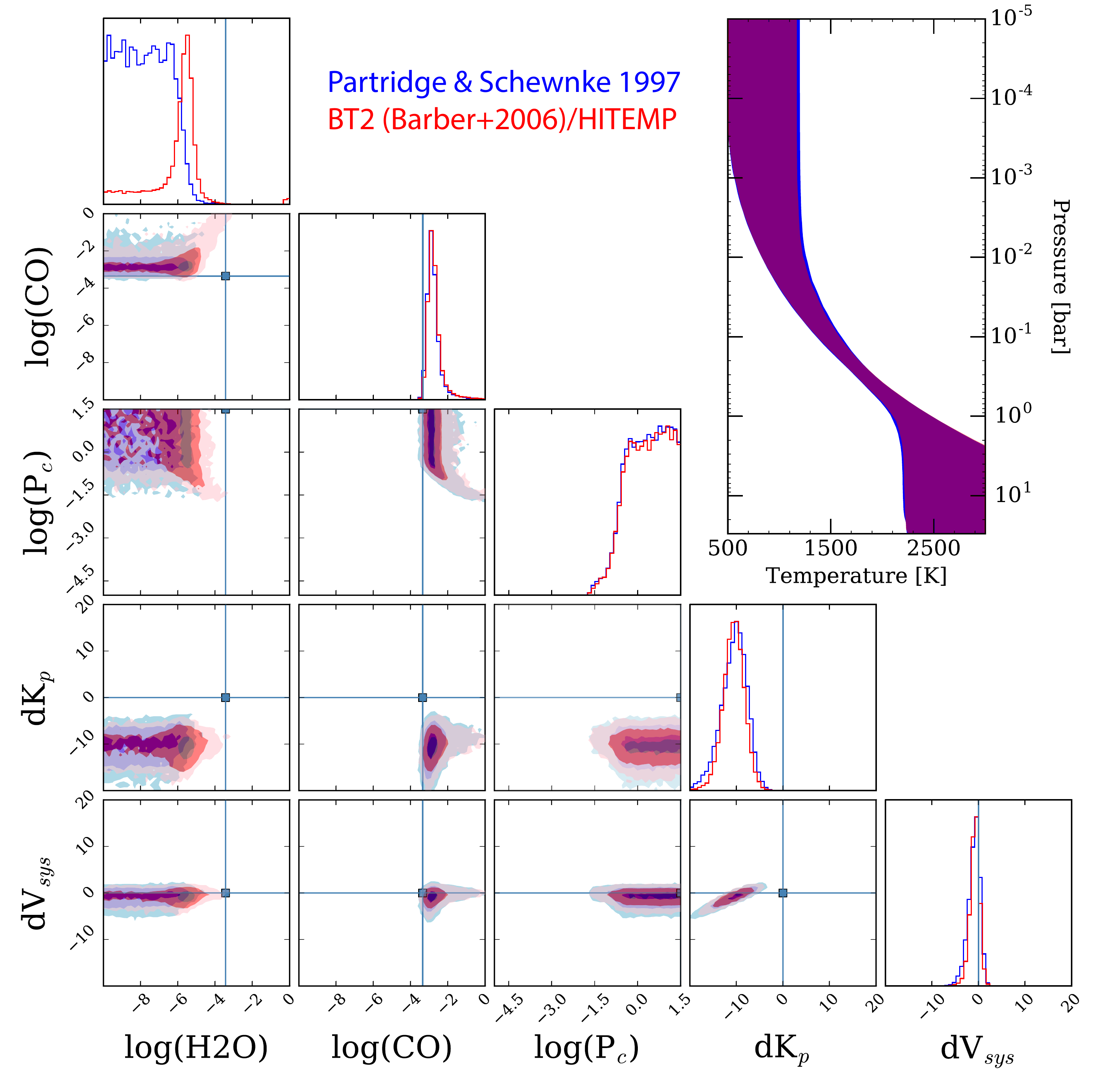}
    \caption{Summary of the HRCCS retrieval results for the CRIRES K-band day-side emission spectrum of HD 209\,458 b.  The blue histograms summarize the posterior under the ``default'' water line list \citep{Freedman2014,PS97}, and red using HITEMP \citep{hitemp2010,Barber06}. Both scenarios assume the \citet{Guillot2010} TP-profile parameterization.  The light blue lines/box within the 1D/2D histograms indicate the approximate mixing ratios predicted by thermo-chemical equilibrium at solar abundance and the 0-offset velocities. An upper limit on the water abundance is obtained under both line-list scenarios, but the CO abundance is fairly tightly constrained at values slightly higher than expected under solar composition, thermochemical equilibrium at these temperatures.}
    \label{fig:HD209_Summary}
\end{figure}

Figure \ref{fig:HD209_Summary} summarizes the posterior probability distribution under the assumption of the default \tp-profile parameterization from Section \ref{sec:retrieval}.  As with HD 189\,733 b we again find only a sub-solar upper limit on the water abundance.  However, we obtain a rather stringent constraint ($\pm$0.3 dex, resulting in a 7.44$\sigma$ evidence based detection) on the CO abundance, of which is only marginally super-solar (0.5 dex higher, or just over 1$\sigma$).  We again find a lower pressure limit on the CTP with a similar, albeit with a less pronounced degeneracy with CO.    

Similar to HD 189\,733 b, the day-side spectrum of HD 209\,458 b was also observed with CRIRES at both 2.3 and 3.2 $\micron$. The 2.3-$\micron$ data were originally published by \citet{schwarz15} and resulted in a marginal detection of CO absorption, no detection of water, and rejection of strong inversion layers. A reanalysis by \citet{bro17} found that the planet signal was very unequally distributed across the four detectors of CRIRES, and could recover CO absorption signal at SNR=5 by unequally weighting the data.

These data along with previously unpublished 3.2-$\micron$ data were recently analyzed by an independent team \citep{hawker18}. 
They report the detection of both \water\ and CO absorption in $K$-band data, and in addition the $L$-band data reveals clear absorption from HCN but not from \water. Therefore, opposite to HD 189\,733 b, water is detected at 2.3 $\micron$ but undetected in the $L$-band.

Contrarily to HD 189\,733 b, though, the CRIRES transmission spectrum of \pname\ at 2.3 $\micron$ does not show any water absorption lines. As this was the first successful detection published with \hires\ \citep{sne10} and the analysis was optimized for CO lines, it is possible that the sensitivity of the data was not sufficient to detect \water.

We again, explore the impact of the water line list, as with HD 189\,733 b.  In general we do not find that the choice of water opacity influences the primary conclusions but it does change the $shape$ of the marginalized water histogram towards a more concentrated solution near 1 ppm. however, there still exists a non-negligible tail extending to lower abundances, suggesting again, that this is still just an upper limit, consistent with a non-detection (according to the Bayesian evidence ratios).  



\subsection{Implications}
It is worth briefly discussing how these results compare to LRS retrievals from emission spectroscopy with HST and Spitzer as both objects have been thoroughly investigated 
with these instruments.  The day side emission spectrum of HD 189\,733 b is by far the most complete in terms of wavelength coverage and ``spectral density'' composed of HST WFC3, NICMOS, Spitzer IRAC/MIPS/IR \citep[see][for a summary]{lin14,Lee2012}, though with questionable reliability of some of the datasets \citep{Gibson2011,Hansen2014}. Retrievals on this emission dataset have suggested water and CO abundances that are broadly consistent with solar thermochemical expectations \citep{lin14,Lee2012}.  Our retrieved abundances for CO are rather consistent with those previous findings \citep[e.g, $\sim10^{-5}-10^{-2}$ in][]{lin14} but our retrieved water abundance upper limit ($\sim 10^{-4}$, Figure \ref{fig:HD189_Summary} top histogram) skirts the lower bounds of previous findings \citep[see, e.g., Table 3 in][]{lin14}.  
The most recent analysis of the fairly complete day-side emission spectrum of HD 209\,458 b \citep{lin16} includes a strong water vapor absorption feature over the HST WFC3 bandpass suggestive of water mixing ratios no lower than $10^{-5}$,  right near the upper limit of what we retrieve (Figure \ref{fig:HD209_Summary} top histogram).  \citet{bro17} investigated both the LRS \citep{lin16} and HRS datasets \citep{schwarz15} and found that water abundances are largely inconsistent due to the fairly stringent lower bound from the LRS data and lack of water detection in the HRS data.

Both objects also have fairly precise HST/WFC3 \citep{dem13,McCullugh13} and HST/STIS \citep{sing11,sing16} transmission spectra with multiple independent analyses suggesting water abundances that span a broad range from $\sim 10^{-6}-10^{-2}$ depending on the specific analysis and datasets used \citep{tsiaras18,LineParm2016,MacMad2017,madhu14}, but no constraints on CO due to the limited wavelength coverage.

The reason for the HRS-LRS water inconsistency is not immediately clear. Whereas one would be tempted to blame it on water line-list uncertainties, our tests reveal no impact of different line lists on the analysis of HD 189\,733 b data, and only a marginal impact on the analysis of HD 209\,458 b. One additional speculation relates to the different atmospheric pressure at which the core of \water\ and CO lines are formed. As low- and high-altitude wind patterns can differ by many \kms\ in hot Jupiters, it is possible that CO and \water\ lines track slightly different Doppler velocity. As water has the biggest dynamic range in terms of weak/strong lines, it is the most affected by vertical wind shears, to the point where the cross-correlation signal is smeared below detectability. Only a comparison with three-dimensional general circulation models as recently shown by \citet{flowers2018} will help us determining if atmospheric circulation has a detectable impact on the day-side spectra of hot Jupiters observed through \hires. In general it would not be surprising that the 3D nature of a planet can and will play a role in interpreting HRS data as these data are observed over a range of orbital phases with the potential for spatially variable temperatures, composition, clouds, and winds.



\section{Conclusions}\label{sec:conclusions}
This work demonstrates that an appropriate mapping of the cross-correlation coefficient to likelihood function allows us to coherently merge hi-resolution cross-correlation spectroscopy with powerful atmospheric retrieval techniques.  This approach opens up a new avenue in interpreting \hires\ data and fully exploits the information buried within it.  Below we summarize our primary developments and findings: 
\begin{itemize}
\item Developed (Section \ref{sec:newlogL}) a novel mapping from cross-correlation to log-likelihood and demonstrated that it is statistically appropriate permitting accurate parameter confidence intervals (Figure \ref{fig:wilks_test}). 
\item Formalized data analysis techniques, in particular the removal of telluric lines, and made them applicable to this new formalism without artificially scaling or biasing the planetary signal (Section \ref{sec:analysis}).
\item Explored the potential for this novel framework to constrain fundamental atmospheric properties like temperatures and abundances on a realistically simulated data-set (Figure \ref{fig:demo_retrieval}, Section \ref{sec:retrieval}). Despite being more degenerate than in \lores\ data, water abundance constraints are comparable to those obtained with HST WFC3. In addition, due to tight constraints on relative molecular abundances of CO and \water, \hires\ + retrievals  permit precise C-to-O ratio determinations.
\item Provided a comprehensive comparison and discussion of the strengths and weaknesses of other correlation-coefficient to log-likelihood mappings (Section \ref{sec:past_map}, \ref{sec:logL_distr}, Figures \ref{fig:wilks_test} and \ref{fig:compare_logL})
\item Determined the significance of the impact that water opacities can have on the results. Precise knowledge of exo-atmosphere relevant opacities are required at hi-resolution for these approaches to work (Section \ref{sec:h2o_xsec}, Figure \ref{fig:linelists}).
\item Explored the impact of missing gases on the retrieved constraints.  Failure to include gases could result in biases and/or artificially broadened constraints (Section \ref{sec:missing_gas}, Figure \ref{fig:missing_gas}).
\item Provided a simple framework for combining HRCCS data and LRS data within a unified likelihood function (Section \ref{sec:lds_hds_combi}, Figure \ref{fig:LDSHDS_retrieval}).  Such an approach leverages the strengths of both types data in a way that is analogous to combining radial velocity and transit data.
\item Applied the framework to existing day-side observations of the transiting hot Jupiters HD 189\,733 b and HD 209\,458 b, obtained with CRIRES at the Very Large Telescope. In spite of a clear signature of CO absorption, \water\ is not clearly detected, regardless of the line list used, \tp\ profiles implemented, or assumptions on the presence of a thick cloud deck. We also rule our confidently the presence of inversion layers in the atmospheres of these two planets.

\end{itemize}

Currently there is a shortage of observations of transiting exoplanets with both HST/WFC3 and high-resolution spectrographs from the ground. The main reason for this is that CRIRES and NIRSPEC, the two most active instruments to provide \hires\ observations in the past 5-10 years, have relatively poor throughput and spectral range. They are thus limited to observe the brightest exoplanets in the sky, which are mostly non-transiting. However, modern spectrographs have drastically superior simultaneous spectral range and equal or better throughput. This increase in sensitivity can be used to move \hires\ observations to smaller telescope facilities \citep{bro18}, to enable observations of fainter planets, or both in case of the most performing instruments such as CARMENES or SPIRou. This technological evolution timely matches the future availability of JWST observations, which will also have increased sensitivity, spectral range, and in some cases spectral resolution. It is also well timed with the recent launch of the TESS satellite, which will find a significant fraction of exoplanets orbiting bright-enough stars to be followed up for atmospheric characterization. Lastly, the age of the next-generation of large telescopes is just a few years ahead. With construction of the Giant Magellan Telescope and the Extremely Large Telescope already ongoing, and first-light high-resolution instrumentation approved for both telescopes, it is crucial to develop techniques to pair the most exquisite JWST observations to the enormous collective power of the GMT and ELT telescopes. 

This paper set the foundation for rigorous analysis of \hires\ datasets and their coherent combination with \lores\ data. Future work will focus on three main directions:
\begin{itemize}
\item Implementing the retrieval on transmission spectroscopy data and jointly analyze all the currently available \lores\ and \hires\ datasets of HD 189\,733 b and HD 209\,458 b as well as other planets.
\item Implementing in the algorithm other telluric removal algorithms such as PCA and Sysrem, which seem to perform excellently at wavelengths affected by strong telluric bands. Currently, the main restriction at implementing these algorithm is purely computational, as the framework forces us to repeat on each tested model spectrum the same exact analysis as on the real data, which comes at a significant computational cost for PCA/Sysrem algorithms.
\item Exploiting the strong predictive power of the framework, in particular for simulating future joint JWST and ELT/GMT/TMT observations of temperate terrestrial worlds. These simulations will identify the optimal wavelength range, spectral resolution, and exposure times to maximize the science return from top-class space and ground-based observatories.
\end{itemize}

\acknowledgments

We thank R. de Kok, I. Snellen, and J. Birkby for the initial discussion on combining high- and low-resolution spectroscopy. We are indebted to the J. Bean and J.-M. D\'esert for in depth discussions leading to the development of this framework. We also thank Ryan MacDonald for permission on adapting his lovely retrieval flow figure for this work.  We would like to thank L. Pino, P. Molli\`{e}re, J. Fortney, and E. Gharib-Nezhad for their unrelenting enthusiasm for us to get this out the door. We would also like to thank Richard Freedman, Roxana Lupu, Mark Marley, and Jim Lyons for invaluable discussions regarding the intricacies of line-lists as well as the incalculable people-hours spent by groups such as HITRAN, ExoMol, and dedicated laboratory scientists on producing top-tier line lists.  We also thank Geoff Blake for taking the time to clarify their team's KECK NIRSpec analysis methods.  We'd also like to thank Ingo Waldmann for a thorough and thoughtful referee report.   Finally, M.R.L acknowledges support from the NASA Exoplanet Research Program award NNX17AB56G and Arizona State University Start Up funds.

%

\vspace{5mm}
\facilities{VLT(CRIRES), HST(WFC3), Spitzer(IRAC)}
\software{python2.7, matplotlib, corner.py, pymultinest/multinest}

\bibliography{biblio}
\bibliographystyle{aasjournal}



\end{document}